\documentclass[a4paper,11pt]{article}
\pdfoutput=1
\usepackage{jheppub}
\usepackage{graphicx}
\usepackage[figuresright]{rotating}
\usepackage{bm,amsmath,amssymb}
\usepackage[mathscr]{eucal}
\usepackage{multirow}
\usepackage{xcolor}
\usepackage{mathrsfs}
\usepackage{mathtools}
\usepackage{slashed}
\usepackage{graphics}
\usepackage{graphicx}
\usepackage{subfigure}
\usepackage{dsfont}
\usepackage{longtable}
\usepackage{bbm} 
\usepackage{float}
\usepackage{tcolorbox}
\usepackage{lipsum}
\usepackage{cancel}
\usepackage{enumerate}
\usepackage{appendix}
\definecolor{lcolor}{rgb}{0.,0.0,0.}
\definecolor{citcolor}{rgb}{0,0.,0.5}

\newcommand{\lo}{{\rm \scriptscriptstyle LO}}

\newcommand{\eqn}[1]{Eq.\,\eqref{#1}}
\newcommand{\beq}{\begin{equation}}
\newcommand{\eeq}{\end{equation}}
\newcommand{\bal}{\begin{align}}
\newcommand{\eal}{\end{align}}
\newcommand{\tr}{{\rm tr}}

\long\def\comment#1{ }

\newcommand{\del}{\partial}

\newcommand{\order}[1]{\mathcal{O}{(#1)}}

\newcommand{\nn}{\nonumber\\}
\newcommand{\rmd}{{\rm d}}
\newcommand{\dif}{{\rm d}}
\newcommand{\et}{\boldsymbol{\epsilon}}
\newcommand{\rme}{{\rm e}}

\newcommand{\bx}{\bm{x}}

\newcommand{\bz}{\bm{z}}

\newcommand{\bb}{\bm{b}}

\newcommand{\bbb}{\overline{\bm{b}}}

\newcommand{\mcal}{\mathcal}

\newcommand{\KT}{K_\perp}
\newcommand{\PT}{P_\perp}
\newcommand{\kg}{k_{g\perp}}

\newcommand{\ellt}{\bm{\ell}}

\newcommand{\Pt}{\boldsymbol{P}}
\newcommand{\Kt}{\boldsymbol{K}}
\newcommand{\ktone}{\boldsymbol{k}_{1}}
\newcommand{\kttwo}{\boldsymbol{k}_{2}}
\newcommand{\kgt}{\boldsymbol{k}_{g}}

\newcommand{\Ccal}{\mathcal{C}}
\newcommand{\Pcal}{\mathcal{P}}

\newcommand{\Ical}{\mathcal{I}}
\newcommand{\Jcal}{\mathcal{J}}

\newcommand{\kt}{\boldsymbol{k}}

\newcommand{\qt}{\boldsymbol{q}}
\newcommand{\Qt}{\boldsymbol{Q}}

\newcommand{\at}{\boldsymbol{a}}
\newcommand{\Bt}{\boldsymbol{B}}

\newcommand{\xt}{\boldsymbol{x}}
\newcommand{\yt}{\boldsymbol{y}}
\newcommand{\zt}{\boldsymbol{z}}

\newcommand{\rt}{\boldsymbol{r}}

\newcommand{\qtone}{\boldsymbol{q_{1}}}
\newcommand{\qttwo}{\boldsymbol{q_{2}}}

\newcommand{\rtone}{\boldsymbol{r_{1}}}
\newcommand{\rttwo}{\boldsymbol{r_{2}}}

\newcommand{\Tr}{{\rm Tr}}

\newcommand{\der}{\mathrm{d}}

\title{Gluon splitting at small $x$: a unified derivation
 for the JIMWLK, DGLAP and CSS equations}
\author[a]{Paul Caucal, }
\author[b]{Edmond Iancu, }
\author[c,d,e]{Farid Salazar, }
\author[f]{Feng Yuan}
 \affiliation[a]{SUBATECH UMR 6457 (IMT Atlantique, Université de Nantes, IN2P3/CNRS), 4 rue Alfred Kastler, 44307 Nantes, France}
 \affiliation[b]{Université Paris-Saclay, CNRS, CEA, Institut de physique théorique, F-91191, Gif-sur-Yvette, France}
\affiliation[c]{Department of Physics, Temple University, Philadelphia, PA 19122 - 1801, USA}
 \affiliation[d]{RIKEN-BNL Research Center, Brookhaven National Laboratory, Upton, New York 11973, USA}
 \affiliation[e]{Physics Department, Brookhaven National Laboratory, Upton, New York 11973, USA}
\affiliation[f]{Nuclear Science Division, Lawrence Berkeley National Laboratory, Berkeley, CA 94720, USA}
 
\abstract{We revisit the calculation of the next-to-leading order (NLO) corrections to dijet production in electron-ion collisions at small $x$.
  We focus on the back-to-back configuration where the relative transverse momentum $P_\perp$ of the measured jets is much larger than
  both their momentum imbalance $K_\perp$ and the target saturation momentum $Q_s(x,A)$. In this regime, we present for the
  first time a complete calculation of the real NLO corrections at leading power in $1/P_\perp$. Our result exhibits
  TMD factorisation, with the same hard factor as at tree-level and a NLO correction to the Weiszäcker-Williams (WW) gluon transverse momentum dependent (TMD) distribution which involves four Wilson-line operators. By studying different kinematical regimes for $K_\perp$ and for the
  radiated gluon, we recover all the quantum evolutions that were previously identified for this process at NLO:
  the B-JIMWLK high-energy evolution and the CSS evolution of the gluon WW TMD, and the DGLAP evolution of the gluon PDF.
  When both $K_\perp$ and the transverse momentum transferred by the target are large compared to $Q_s$, all the Wilson-line
  operators boil down to the unintegrated gluon distribution and our NLO result for the gluon TMD can be used to isolate the
  transverse-momentum dependent gluon splitting function.}

\begin{document}
\maketitle
\newpage 

\section{Introduction}

A major current challenge in QCD and hadron physics is the characterisation of the partonic structure of the proton and nucleons bounded inside heavy nuclei in terms of three-dimensional, transverse momentum dependent parton distribution functions (TMDs), which depend not only on the longitudinal momentum fraction $x$ of the hadron carried by the partons but also on their transverse momentum $k_\perp$~\cite{Boussarie:2023izj}. Experimentally probing the TMDs, especially the gluon TMD~\cite{Arleo:2025oos}, is a key scientific goal of the Electron Ion Collider (EIC) that is set to be built at the Brookhaven National Laboratory in the United States~\cite{AbdulKhalek:2021gbh}. In the regime where $x\ll 1$ (typically, $x\lesssim 10^{-2}$), the partonic content of hadrons is dominated by gluons, and the effective theory of the Colour Glass Condensate (CGC)~\cite{Iancu:2002xk,Iancu:2003xm,Gelis:2010nm,Kovchegov:2012mbw,Morreale:2021pnn} provides first principle determination of the gluon TMDs~\cite{Dominguez:2011wm,Metz:2011wb,Dominguez:2011br,Marquet:2017xwy,Iancu:2021rup,Iancu:2022lcw,Iancu:2023lel,Altinoluk:2024tyx}, as well as those of sea quarks~\cite{Marquet:2009ca,Xiao:2017yya,Hatta:2022lzj,Hauksson:2024bvv,Caucal:2025xxh}. This CGC/TMD correspondence, first elucidated in the seminal paper \cite{Dominguez:2011wm}, is particularly remarkable and useful for constraining small-$x$ TMDs without relying on the conventional strategies at  moderate $x$, which involve either global fits of key observables (semi-inclusive Deep Inelastic Scattering and Drell-Yan lepton-pair production in proton-proton collisions)~\cite{Bacchetta:2024qre,Bacchetta:2025ara,Moos:2025sal}, or lattice QCD calculations~\cite{Constantinou:2020pek}, to constrain the non-perturbative part of the TMDs~\cite{Shanahan:2020zxr,Avkhadiev:2023poz,Avkhadiev:2024mgd,Bollweg:2025iol}.

Recently, the CGC/TMD correspondence has been the subject of extensive studies, which can schematically be divided into three directions: \texttt{(i)} this correspondence has been extended to additional processes~\cite{Altinoluk:2018byz,Altinoluk:2020qet,Tong:2022zwp,Tong:2023bus} including in particular particle production in diffractive processes~\cite{Iancu:2021rup,Iancu:2022lcw,Iancu:2023lel,Hatta:2022lzj,Hauksson:2024bvv} and in the target fragmentation region~\cite{Caucal:2025qjg}; \texttt{(ii)} sub-eikonal~\cite{Altinoluk:2024tyx,Altinoluk:2024zom,Agostini:2024xqs,Altinoluk:2023qfr,Altinoluk:2022jkk} and kinematic power corrections~\cite{Altinoluk:2019fui,Altinoluk:2019wyu,Mantysaari:2019hkq,Boussarie:2021ybe} to tree-level TMD factorised cross-sections have been investigated at small $x$; \texttt{(iii)} the connection between CGC and TMD has been generalised beyond tree level~\cite{Mueller:2012uf,Mueller:2013wwa,Taels:2022tza,Caucal:2022ulg,Caucal:2024bae,Caucal:2024vbv,Duan:2024nlr,Duan:2024qev} by consistently incorporating various quantum evolution effects that resum large logarithms --- namely, small-$x$ (or high-energy) evolution from the Balitsky-Kovchegov (BK) or Jalilian-Marian-Iancu-McLerran-Weigert-Leonidov-Kovner~(JIMWLK) equation~\cite{Balitsky:1995ub,Kovchegov:1999yj,JalilianMarian:1997jx,JalilianMarian:1997gr,Kovner:2000pt,Weigert:2000gi,Iancu:2000hn,Iancu:2001ad,Ferreiro:2001qy}, Colllins-Soper-Sterman (CSS) evolution~\cite{Collins:1981uk,Collins:1981uw,Collins:1984kg,Collins:2011zzd}, and Dokshitzer-Gribov-Lipatov-Altarelli-Parisi (DGLAP) evolution~\cite{Gribov:1972ri,Altarelli:1977zs,Dokshitzer:1977sg} --- as well as by computing the hard factor for key processes at next-to-leading order (NLO)~\cite{Caucal:2023nci,Caucal:2023fsf}.

\begin{figure}
    \centering
    \includegraphics[width=0.8\linewidth]{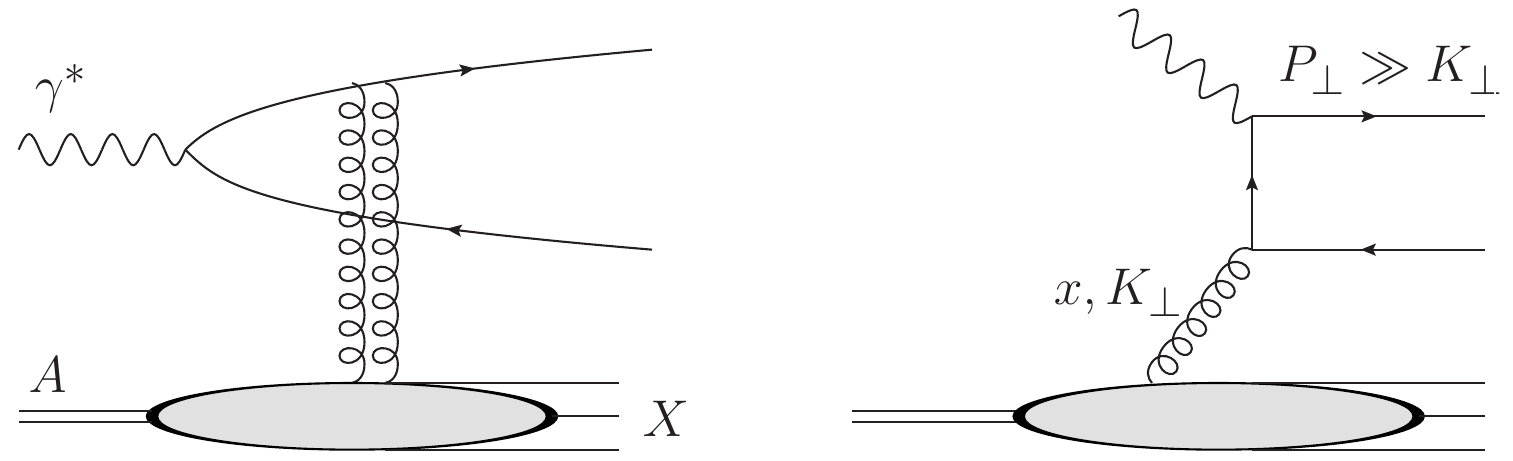}
    \caption{(Left) CGC dipole picture for the $\gamma^*$-nucleus reaction producing two jets at leading order in the colour dipole frame and including multiple gluon scattering. (Right) Target picture of the same process in the conventional TMD factorisation approach. The $t$-channel gluon knocked out of the collision has intrinsic transverse momentum $K_\perp$, much smaller than the transverse momentum $\sim P_\perp$ of each individual jet produced in the final state.}
    \label{fig:s-vs-t-channel-picture}
\end{figure}

This article follows up on this third point. We revisit the production of a back-to-back dijet pair in electron-nucleus ($eA$) deep inelastic scattering (DIS) at NLO in the CGC, focusing on certain real contributions that were previously overlooked in earlier studies of this process at NLO (essentially because they do not contribute to the various quantum evolutions of the gluon TMD, i.e. they are not enhanced by large, rapidity or transverse, logarithms).
To leading order (LO) in the CGC effective field theory (EFT), this process admits TMD factorisation in terms of the Weiszäcker-Williams (WW) gluon TMD in the limit where the dijet relative transverse momentum $P_\perp$ is much larger than both the dijet transverse momentum imbalance $K_\perp$ and the nucleus saturation scale $Q_s$. Such a factorisation is far from being obvious in the CGC EFT supplemented by the colour dipole picture~\cite{Kopeliovich:1981pz,Bertsch:1981py,Mueller:1989st,Nikolaev:1990ja} for high energy $\gamma^*$-nucleus interaction. Indeed, in this picture, the virtual photon with large light-cone momentum $q^+\gg Q$ (with $Q^2$ the photon virtuality, which for simplicity is taken to be of the same order as the hard scale $P_\perp^2$ in this introduction) splits into a quark-antiquark ($q\bar q$) pair which then interacts with the shock-wave background field of the dense target and eventually becomes the dijet pair measured in the final state. As first shown in~\cite{Dominguez:2011wm}, to leading power in $1/P_\perp$ this dipole picture matches the more conventional target picture of TMD factorisation (see e.g.~\cite{delCastillo:2020omr,delCastillo:2021znl} as well as  Fig.~\ref{fig:s-vs-t-channel-picture}), where the process  factorises between a hard factor describing the partonic process $\gamma^*g\to q\bar q$ and the WW gluon TMD expressing the probability density of finding the incoming gluon in the target with longitudinal momentum fraction $x$ (w.r.t. the target) and transverse momentum $K_\perp$.

\begin{figure}
    \centering
    \includegraphics[width=0.8\linewidth]{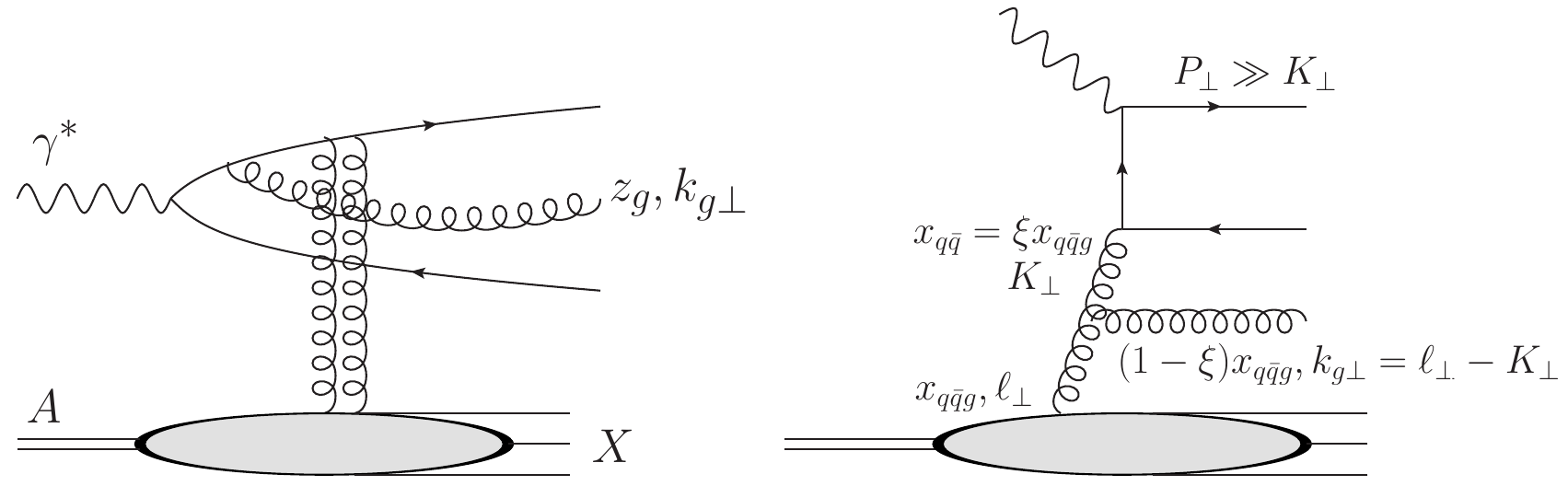}
    \caption{CGC (left) vs target (right) pictures of the real NLO corrections (to be compared with Fig.~\ref{fig:s-vs-t-channel-picture}). Depending on the kinematics $(z_g,k_{g\perp})$ of the radiated gluon, it contributes either to the high-energy or CSS evolution of the WW gluon TMD, or to the $\mathcal{O}(\alpha_s)$ finite corrections (without logarithms) to the WW gluon TMD.}
    \label{fig:s-vs-t-channel-picture-nlo}
\end{figure}

Beyond leading order (LO), the study of TMD factorisation at small $x$ is complicated by the fact that, within the CGC approach, the NLO corrections are computed as gluon emissions by the {\it dilute projectile} (here, the $q\bar q$ color dipole), whereas the TMD factorisation is truly a {\it target} picture --- indeed, the WW gluon TMD refers to small-$x$ gluons from the dense nucleus. Hence, in order for this factorisation to be preserved by the NLO corrections (as computed to leading power in $1/P_\perp$), one has to be able to reinterpret these corrections as renormalisations of the hard factor and of the TMD. In particular, one must recover the evolution equations which describe the quantum evolution of the WW gluon TMD with decreasing $x$ and/or increasing $P_\perp^2$. This ultimately means that some of the effects of the gluon emissions by the color dipole should be equivalent to parton splittings occurring within the target wavefunction
(see Fig.~\ref{fig:s-vs-t-channel-picture-nlo} for a graphical illustration). Since moreover the target is dense, these splittings should be affected by gluon saturation. The fact that this strategy works in practice has been already demonstrated for various processes~\cite{Iancu:2022lcw,Hauksson:2024bvv,Caucal:2024bae,Caucal:2024vbv} and will be further confirmed by our results in this paper.


The evolution of the WW gluon TMD with decreasing $x$ can be obtained by acting with the JIMWLK Hamiltonian on the corresponding operator, without explicit reference to the underlying process. This procedure leads to a special version of the B-JIMWLK equations, known as the Dominguez-Marquet-Munier-Xiao (DMMX) equation~\cite{Dominguez:2011gc}. This is a non-linear and also non-closed equation (it relates the evolution of the WW gluon distribution to other gauge-invariant multi-gluon correlators built with Wilson lines), whose solution resums perturbative corrections enhanced by the rapidity logarithm $\ln(1/x)$. More recently, it has been explicitly verified~\cite{Taels:2022tza,Caucal:2022ulg} that the DMMX equation is indeed generated by the NLO corrections to dijet production and, more precisely, by the gluon emissions which are sufficiently soft (in a sense to be shortly specified) and which occur relatively close to the scattering off the nuclear shock-wave (either before, or after, this scattering). But the hierarchy of transverse momentum scales in the problem ($P_\perp\gg K_\perp, Q_s$) also introduces NLO corrections enhanced by large transverse logarithms, like $\ln(P_\perp^2/K_\perp^2)$. In~\cite{Mueller:2013wwa,Taels:2022tza,Caucal:2022ulg}, the leading contributions of this type, the so-called Sudakov double and single logarithms of the ratio $P_\perp^2/K_\perp^2$, have been isolated at small $x$ and shown to be consistent with TMD factorisation: they can be absorbed into a renormalisation of the gluon TMD, as usually done at moderate $x$. In~\cite{Caucal:2023nci,Caucal:2023fsf}, this construction has been extended to the finite NLO corrections arising from virtual gluon emissions and from real gluon emissions occurring in the final state (after the collision with the shock-wave).
However, a remaining class of real corrections has been ignored, those where the gluon is emitted before the scattering. As shown in~\cite{Caucal:2024bae}, these diagrams too contribute to the leading power and should therefore be computed and included among the NLO corrections to the cross-section for inclusive back-to-back dijet production in DIS. In Ref.~\cite{Caucal:2024bae}, their contribution has been computed under the assumption that the gluon transverse momentum $\kg$ is comparable with the dijet imbalance $K_\perp$ and they are both much larger than 
the target saturation momentum ($\kg\simeq K_\perp\gg Q_s$) --- that is, the dijet  imbalance is controlled by the gluon recoil. With this kinematics, gluon emissions before and after scattering combine with each other to generate the DGLAP evolution of the gluon PDF together with a particular, double-logarithmic, approximation to the CSS evolution of the gluon TMD~\cite{Caucal:2024bae}.

In this paper, we will generalise these previous results by presenting for the first time the complete results for the real NLO corrections to  back-to-back 
dijet production in $eA$ DIS, to leading-power in $1/P_\perp$ and for generic values of the dijet imbalance $K_\perp$ (with $K_\perp \ll P_\perp$, of course), including semi-hard values $K_\perp \sim Q_s$. This in turn requires the calculation of gluon emissions that can occur either before or after scattering, which carry generic transverse momenta $\kg\ll P_\perp$ and also generic values for the longitudinal momentum fraction $z_g\equiv k_g^+/q^+$ (with $z_g\lesssim\kg /P_\perp\ll 1$ though, for reasons to be explained shortly). This goes beyond the previous derivations of the DMMX equation in Refs.~\cite{Taels:2022tza,Caucal:2022ulg}, which were restricted to much softer gluons with $z_g\ll \kg^2 /P_\perp^2$ (see below). It also goes beyond the calculations of the Sudakov effects in Refs.~\cite{Mueller:2013wwa,Taels:2022tza,Caucal:2022ulg} and their extensions in~\cite{Caucal:2023nci,Caucal:2023fsf}, which were restricted to final-state emissions. It finally goes beyond the recent developments in Ref.~\cite{Caucal:2024bae} by enforcing transverse-momentum conservation and by allowing $K_\perp$ and $\kg$ to also take smaller values, of order $Q_s$, for which saturation effects become important. Given the generality of our calculations, we will recover (the real-gluon contributions to) all the evolution equations previously established for the gluon WW TMD and the gluon PDF --- that is, DMMX, DGLAP and CSS. We will moreover extend the CSS equation to single-logarithmic accuracy, as standard at moderate $x$. Last but not least, we will uncover additional leading-power contributions, which are pure $\order{\alpha_s}$ effects ----- that is, they are not accompanied by large kinematical logarithms, hence do not contribute to any evolution. Importantly, all these NLO corrections preserve TMD factorisation with the same hard factor as at tree-level: all the leading-power, real-gluon, NLO corrections can be absorbed into the renormalisation of the WW gluon TMD.

To disentangle the various types of quantum evolution, it is important to properly identify the contributing regions in the three-dimensional phase-space $(z_g, \kgt)$ for gluon emissions. Since we are interested in jet production, we can ignore real gluon emissions at small angles, inside the cones defining the two measured jets (their effects would anyway largely cancel against the corresponding virtual emissions.
Hence, we only need to consider real gluons emissions at large angles, away from the two jets, which are moreover compatible with the back-to-back configuration of the quark-antiquark pair. Clearly, this condition excludes the emissions where the transverse momentum $\kg$ of the radiated gluon is parametrically larger than the dijet imbalance $K_\perp$ (such hard emissions
would modify the structure of the final state). It also excludes the emissions with {\it too} low values of $\kg$, but only so long as the respective emission angles, estimated as $\theta_g\sim \kg/(z_g q^+)$, remain smaller than the jet opening angle, which is parametrically of the order $\theta_q\sim P_\perp/q^+ $ (say, for the quark jet). The allowed emissions must obey $\theta_g\gtrsim \theta_q$, or $z_g\lesssim \kg/P_\perp$, which together with $\kg \lesssim K_\perp$ implies that the interesting gluons are {\it soft}:  $z_g \lesssim K_\perp/P_\perp\ll 1$. But not all such soft gluons contribute to the high-energy evolution: as mentioned earlier, the ``genuinely small-$x$'' gluons (those which contribute to the B-JIMWLK evolution) are those whose formation times $\Delta x^+_g\sim 2z_g q^+/\kg^2$ are (much) smaller than the coherence time $\Delta x^+_{\gamma^*}\sim 2q^+/Q^2$ of the virtual photon. For the interesting situation where $Q^2\sim P_\perp^2$, this argument implies that only the {\it very} soft gluons with $z_g\ll \kg^2/P_\perp^2$ should be associated with the high-energy evolution. As we shall see in what follows, the logarithmic phase space at $x \kg^2/P_\perp^2 \ll z_g \ll \kg^2/P_\perp^2$ (where $x\sim Q^2/s\ll 1$ is the Bjorken variable) generates the (real part of the) DMMX equation for the evolution of the WW gluon TMD with decreasing $x$.

The remaining phase space for soft gluon emission, within the range $\kg^2/P_\perp^2 \lesssim z_g\lesssim \kg/P_\perp$, turns out to be more subtle.  The logarithmic contribution from this region fully comes from final state gluon emissions in the dipole picture and builds the real part of the CSS equation. The solutions to the equation 
achieve the resummation of the Sudakov logarithms within the structure of the gluon TMD. As shown in~\cite{Taels:2022tza,Caucal:2022ulg,Caucal:2023fsf}, the proper separation between this logarithmic phase-space and that for the high-energy evolution, as occurring at $z_g \sim \kg^2/P_\perp^2$, is essential in order to obtain the correct expressions for the Sudakov logarithms. This also ensures the proper matching between the NLO impact factor and the collinearly-improved version of the B-JIMWLK evolution~\cite{Beuf:2014uia,Iancu:2015vea,Ducloue:2019ezk,Boussarie:2025mzh}, which solves the instability problems occurring at NLO~\cite{Balitsky:2008zza,Lappi:2015fma,Lappi:2016fmu}. Moreover, gluon emissions with $z_g$ values around the borderline, $z_g \sim \kg^2/P_\perp^2$, play an important role by themselves: as shown in~\cite{Caucal:2024bae}, they generate the DGLAP evolution of the gluon PDF. Ref.~\cite{Caucal:2024bae} focused on the special regime where the dijet imbalance is caused by the gluon recoil ($\kg\simeq K_\perp\gg Q_s$), which is responsible for both the DGLAP evolution and the dominant, double-logarithmic, approximation to the CSS equation. In what follows, we will relax this approximation ($\kg$ and $K_\perp$ can take generic values), which will allow us to obtain the full CSS equation with transverse momentum conservation. 

So far, our discussion has mostly focused on the NLO corrections which are responsible for the quantum evolutions of the WW gluon TMD. However, there are additional, real NLO corrections, which matter to leading power as well, but are not enhanced by kinematical logarithms (so, they do not matter for the evolution equations). These additional contributions, which are associated with initial-state emissions in the dipole picture, will be computed here for the first time. They are under control in our calculation because we allow the transverse momenta $K_\perp$ and $\kg$ to take generic values and we treat exactly the transverse momentum dependence of the collision between the gluon and the shock-wave.  When $K_\perp$ and $\kg$ are comparable to $Q_s$, saturation effects are important and are encoded in the same Wilson-lines operators as appearing in the DMMX equation for the WW gluon TMD. In the linear/dilute regime where $K_\perp$ and $\kg$ are large compared to $Q_s$ (meaning that the transverse momentum $\ell_\perp$ transferred by the target is large as well, $\ell_\perp\gg Q_s$), all the aforementioned Wilson-line operators (in particular, the WW gluon TMD) reduce to the unintegrated gluon distribution (UGD) describing small-$x$ gluons in the target in the absence of saturation. Then the ensemble of the leading-power NLO corrections to the UGD can be interpreted as describing a splitting $g\to gg$ which occurs inside the target wavefunction with a transverse-momentum dependent splitting function (see the right diagram in Fig.~\ref{fig:s-vs-t-channel-picture-nlo}). Such a splitting function can be useful if one aims at solving the evolution equation for the UGD using Monte-Carlo techniques~\cite{Hautmann:2022xuc}.

Our paper is divided as follows. In Sect.~\ref{sec2}, we provide a brief overview of the LO result for the inclusive back-to-back dijet cross-section in the color dipole picture at small $x$ and of the CGC/TMD correspondence for this specific process. In Sect.~\ref{sec:WW-splitting}, we construct the real NLO cross-section at full leading power in the simultaneous expansion in powers of $K_\perp^2/P_\perp^2$ and $Q_s^2/P_\perp^2$. To that aim, we compute the leading power contributions generated by real gluon emissions with generic transverse momenta $\kg \ll P_\perp$ and with longitudinal momentum fraction $z_g$ (w.r.t. the photon) within the range $x\kg^2/P_\perp^2 \lesssim z_g\lesssim \kg/P_\perp$. In Sect.~\ref{sec4}, we first explain how to isolate from this result the large kinematical logarithms which appear in the dijet cross-section at NLO, namely the small-$x$ logarithm $\ln(1/x)$, the Sudakov double logarithm $\ln^2(P_\perp/K_\perp)$, and the DGLAP logarithm $\ln(Q^2/Q_s^2)$. Then we use these logarithmically enhanced NLO contributions to deduce the real contributions to the various evolution equations satisfied by the WW gluon TMD: the DMMX equation (a special equation from the B-JIMWLK hierarchy), the CSS equation, and the DGLAP equation for the gluon PDF (as obtained from the TMD). In Sect.~\ref{sec5}, we present a summary of our results (notably, an equation which exhibits the general structure of TMD factorisation for the back-to-back dijet cross-section at NLO and in the non-linear regime at small $x$) together with an outlook for future work.

\section{Back-to-back dijet production at leading order}
\label{sec2}

We start with a short review of the leading-order (LO) result for the cross-section for back-to-back dijet production in DIS at small $x$ (a comprehensive discussion, including details on the derivation, can be found in \cite{Dominguez:2011wm}). This is useful to fix the notations and also in view of the comparison with the next-to-leading order (NLO) results to be obtained later. As usual with the DIS calculations at small $x$, we work in the dipole frame where the virtual photon $\gamma_\lambda^*$ with four-momentum $q^\mu=(q^+,q^-= - Q^2/2q^+,\boldsymbol{0})$ (in light cone coordinates) is a relativistic right mover with $q^+\gg Q$ and polarisation $\lambda$. In this frame, the photon first splits into a quark-antiquark pair (the color dipole) which then interacts with the gluon field of the nuclear target, represented by a classical shock-wave. The nucleus too is ultra-relativistic, with four momentum $P_N^\mu=(0,P_N^-,\boldsymbol{0})$ per nucleon (we systematically neglect the nucleon mass, assumed to be small compared to $P_N^-$). The four-momenta of the quark and the antiquark are noted $k_1^\mu$ and $k_2^\mu$ respectively.

We shall focus on the case of a transversely polarised virtual photon; analogous results can be obtained for a virtual photon with longitudinal polarisation. At LO in perturbation theory, the two fermions can be assumed to be on-shell after the collision and identified with the final jets.  We note $\Pt$ and $\Kt$ are the relative transverse momentum and the transverse momentum imbalance of the two measured jets, with
\begin{align}
    \Pt=z_2\ktone-z_1\kttwo\,,\quad \Kt=\ktone+\kttwo\,,
    \label{defPK}
\end{align}
and $z_i=k_i^+/q^+$ the photon's longitudinal momentum fractions carried by the jets, which obey $z_1+z_2=1$. (Indeed, in the eikonal approximation at high energy, the longitudinal momenta of the two fermions are not modified by the collision.)

At LO, and up to power corrections in $K_\perp^2/P_\perp^2$ and $ Q_s^2/P_\perp^2$, the differential cross-section for the $\gamma^*_T+A\to q\bar q+X$ process reads
\begin{align}
	\label{sigma0}
	\frac{\dif \sigma^{\gamma_{T}^* A \to q\bar{q}X}_\lo}
	{\rmd z_1
  	\rmd z_2\dif^2\Pt\dif^2\Kt } &\,=   \mathcal{H}_{\rm LO}^{mn}(\Pt, Q,z_1,z_2)\mcal{W}^{mn,(0)}(x,\Kt)\\
    &={\alpha_{em}\alpha_s}
	e_f^2\delta(1-z_1-z_2)\left(z_1^{2} + 
	z_2^{2}\right) \frac{P_{\perp}^4 + \bar{Q}^4}
	{(P_{\perp}^2 + \bar{Q}^2)^4}\nonumber\\*[0.2cm]
	&\, \times \left\{\,\mcal{F}_g^{WW,(0)}(x, K_{\perp})
	 -   \frac{4 P_{\perp}^2 \bar{Q}^2}{P_{\perp}^4 + \bar{Q}^4}
        \left[ \frac{( \Pt\cdot \Kt)^2}{P_{\perp}^2\,K_{\perp}^2 }
       -\frac{1}{2}
        \right]  \mcal{H}_g^{WW,(0)}(x,K_{\perp})\right\}\,.
\end{align}
The LO hard factor for the $\gamma^*_TA\to q\bar q$ process is defined as
\begin{align}\label{HLO}
    \mathcal{H}_{\rm LO}^{mn}(\Pt, Q,z_1,z_2)&\equiv\alpha_{em}\alpha_s e_f^2\delta(1-z_1-z_2)\frac{z_1^2+z_2^2}{(P_\perp^2+\bar Q^2)^2} \left[\delta^{mn}-\frac{4\bar Q^2\Pt^m\Pt^n}{(\Pt^2+\bar Q^2)^2}\right] \,,
\end{align}
where $\alpha_{em}$ is the electromagnetic coupling constant, $e_f^2$ is the sum of the squared of the light quark charges, and $\bar{Q}^2 = z_1 z_2 Q^2$. The distributions $\mcal{F}_g^{(0)}(x, K_{\perp})$ and $\mcal{H}_g^{(0)}(x,K_{\perp})$ are the (leading-order) unpolarised and respectively 
linearly polarised Weiszäcker-Williams (WW) gluon TMDs~\cite{Metz:2011wb}. They are respectively defined as the trace and the traceless part of the 2-point function of the transverse components of the target field $\mcal{A}^m(\bb)=\frac{i}{g}V(\bb)\partial^m V^\dagger(\bb)$, with $m=1,2$
and $V(\bb)$ a light like Wilson line in the fundamental representation of $SU(3)$ describing the color precession of a quark with transverse coordinate $\bb$:
\begin{align}
    V(\bb) = \Pcal \left[\exp\left(ig \int_{-\infty}^{\infty} \der x^+ A_a^-(x^+,\bb) t^a \right) \right] \,, 
    \label{eq:Wilson-line}
\end{align}
where $\Pcal$ denotes path ordering on the exponential of the colour matrix. Then, the tensorial WW gluon TMD reads
\begin{align}
 \hspace*{-0.2cm}
 	\label{dGWW} \mcal{W}^{mn,(0)}(x, \Kt)
	&\ \equiv  \,
  \int\frac{\rmd^2{\bm{b}}\, \rmd^2 \bbb}{(2\pi)^4} \ 
   \rme^{-i\Kt\cdot(\bb-\bbb)}\   \frac{-2}{\alpha_s}
\left\langle \! {\rm tr} \!\left[ 
V_{\bb} (\del^m V_{\bb}^{\dagger})\,
     V_{\bbb} (\del^n V_{\bbb}^{\dagger})\right]\right\rangle_{x}
     \nonumber\\*[0.2cm]
	&\ \equiv \,\frac{\delta^{mn}}{2}\,
\mcal{F}_g^{WW,(0)}(x, K_\perp) +
        \left(\frac{\Kt^m \Kt^n}{K_{\perp}^2}
        -\frac{\delta^{mn}}{2} \right)
        \mcal{H}_g^{WW,(0)}(x,K_\perp)\,.
    \end{align}
The brackets $\langle ... \rangle_x$ denote the CGC average over the target gluon field $A_a^-$. The variable $x$ represents the fraction of the target longitudinal momentum $P_N^-$ that is probed by the collision. Specifically, the CGC average must be evaluated over the gluon ensemble created by the high energy evolution {\it of the nuclear target}, from some initial value $x_0\sim 10^{-2}$ (where the small-$x$ approximations start to be appropriate) down to final value $x\ll x_0$ of interest. This recipe for evaluating the high energy evolution may look natural from a physics standpoint, yet it is not obviously realised in the CGC EFT, where quantum corrections are rather computed as gluon emissions by the {\it dilute projectile} --- here the $q\bar q$ dipole (see the next section). That is, the rapidity variable which {\it a priori} plays the role of the ``evolution time'' for the B-JIMWLK evolution is the ``plus'' rapidity difference w.r.t. the projectile, and not the ``minus'' rapidity difference w.r.t. the target. In the presence of largely separated transverse momentum scales, the respective phase-spaces can be quite different. This mismatch leads to large higher-order corrections and related instabilities in the BK/JIMWLK equations at NLO and beyond. Remarkably, the dominant such corrections can be effectively resummed to all orders by merely replacing the projectile rapidity by the target rapidity as the evolution variable in the LO equations~\cite{Beuf:2014uia,Iancu:2015vea,Ducloue:2019ezk,Boussarie:2025mzh}. As we shall see, a careful consideration of the phase-space for gluon emissions by the color dipole directly leads to BK/JIMWLK equations for which the evolution time is the target rapidity. This justifies our choice of the target longitudinal momentum fraction $x$ as the rapidity parameter for the CGC average in \eqn{dGWW}.

At LO, the relevant value of $x$ is the fraction  $x_{q\bar q}$ of the target longitudinal momentum that is transferred to the $q\bar q$ pair, in order for the latter to emerge on-shell after the collision, as determined by the condition of minus longitudinal momentum conservation:
\begin{align}\label{xqqdef}
    x_{q\bar q}\equiv \frac{1}{2q^+P_N^-}\left[Q^2+\frac{k_{1\perp}^2}{z_1}+\frac{k_{2\perp}^2}{z_2}\right]\simeq \frac{Q^2+M_{q\bar q}^2}{W^2+Q^2}\,,
\end{align}
where the approximate equality holds in the back-to-back limit. Here, we have introduced the dijet invariant mass $M^2_{q\bar q}\equiv (k_1+k_2)^2\simeq P_\perp^2/(z_1z_2)$ and $W^2\equiv (q+P_N)^2\simeq 2q^+P_N^--Q^2$ up to target mass corrections.

Notice finally that the structure of the hard factor \eqref{HLO} favors the production of dijets with $P_\perp^2 \sim \bar Q^2$ (or $M_{q\bar q}^2\sim Q^2$). To avoid a proliferation of scales, throughout our analysis we assume that $P_\perp$ and $Q$ are both hard and comparable with each other. Accordingly,  the typical jets that we consider are rather symmetric: $z_1\sim z_2\sim 1/2$.

\section{Real NLO corrections to the Weiszäcker-Williams gluon TMD}
\label{sec:WW-splitting}

We now wish to investigate the real NLO corrections to the cross-section for the production of a quark-antiquark dijet, with the two jets propagating back-to-back in the transverse plane. In the color dipole picture, these corrections are associated with the emission of a gluon by either the quark, or the antiquark, leg of the dipole. We use $z_g$ to denote the longitudinal momentum fraction of the gluon w.r.t. the incoming photon, $z_g\equiv k_g^+/q^+$,  and $\kgt$ for its transverse momentum. The dijet relative ($\Pt$) and total ($\Kt$) transverse momenta are defined as at LO, cf. \eqn{defPK}, and satisfy $\PT\gg \KT$. We also introduce the total transverse momentum transferred from the target $\ellt$, defined as
\begin{align}\label{defl}
    \ellt&=\ktone+\kttwo+\kgt = \Kt + \kgt \,.
\end{align}
Since we are interested in back-to-back dijets, we shall consider the regime where the emitted gluon is relatively soft, $z_g\lesssim \kg /P_\perp$ (and thus propagates at large angles), while the dijet relative momentum is the largest 
transverse momentum scale in the problem, 
$P_\perp\gg K_\perp,\ell_\perp, \kg, Q_s$ without any specific hierarchy between $K_\perp$, $\kg$, $\ell_\perp$, and $Q_s$. 
Within this regime, we shall focus on NLO corrections associated with real gluon emissions (i.e. radiative processes where the emitted gluon emerges in the final state, although it is not measured in the detectors) and we shall extract the leading-power contributions to these corrections --- i.e. the leading order terms in the simultaneous expansion in powers of $K_\perp^2/P_\perp^2$ and $Q_s^2/P_\perp^2$ (with typically $\ell_\perp\lesssim \KT$). As we shall see, the complete result to the accuracy of interest involves gluon emissions occurring both before (``initial-state'') and after (``final-state'') the collision with the nuclear shock-wave.

\subsection{Real gluon emission at leading power: the amplitude}

For detailed derivations of the $q\bar qg$ component of a light-cone wave-function (LCWF) or a scattering amplitude, we refer the reader to~\cite{Beuf:2011xd,Ayala:2017rmh,Iancu:2018hwa,Caucal:2021ent,Iancu:2022gpw,
Bergabo:2023wed}. Here we will rather follow the economical approach of~\cite{Caucal:2024bae}: we start with the general structure of the LCWF in transverse momentum space, as obtained by straightforwardly applying the Feynman rules of light-cone perturbation theory, and simplify this expression by using our kinematical assumptions $P_\perp\gg K_\perp, \kg,\ell_\perp,  Q_s$ and $z_g\lesssim \kg/P_\perp$. An independent derivation using covariant perturbation theory is presented in Appendix~\ref{app:amplitude-qqbarg}, including the case where the virtual photon is longitudinally polarised. The only difference with respect to~\cite{Caucal:2024bae} is that here we will relax the assumption $\KT\simeq \kg\gg \ell_\perp, Q_s$ (vectorially, $\Kt \simeq -\kgt$),  which is the statement  that the dijet imbalance is hard and controlled by the recoil of the gluon emission. In practice, this assumption was used in~\cite{Caucal:2024bae} merely to simplify the structure of the $S$-matrix, that is, the colour structure of the amplitude for gluon emission. For this reason, we can simply take over the structure of the emission vertex from Ref.~\cite{Caucal:2024bae} and only highlight the differences which appear in the colour structure when allowing $\KT$ and $\kg$ to take arbitrary values, including semi-hard values of order $Q_s$.

\paragraph{Gluon emission before the shock-wave.} For gluon emissions in the initial state, the colour structure of the amplitude is a priori $U^{ac}_{\zt}V_{\xt}t^{c}V^{\dagger}_{\yt}-t^a$ where $\xt,\yt$ and $\zt$ are respectively the transverse coordinates of the quark, antiquark and gluon, $V_{\xt}$ and $V^{\dagger}_{\yt}$ are Wilson lines in the fundamental representation, cf. \eqn{eq:Wilson-line}, while $U_{\zt}$ is a Wilson line in the adjoint representation, as obtained by replacing $t^t\to T^a$ in  \eqn{eq:Wilson-line}. This $S$-matrix reduces to that of an effective gluon-gluon dipole, made with the physical gluon at $\zt$ and with the small $q\bar q$ pair, with transverse size $r=|\xt-\yt|\sim 1/\Pt$ and central position  $\bb=z_1\xt+z_2\yt$: after emitting the gluon, this pair remains in a color octet state and thus behaves like a gluon at $\bb$. Indeed, we can write
\begin{align}
\label{Uin}
	U^{ac}_{\zt}V_{\xt}t^{c}V^{\dagger}_{\yt}-t^a&\,\simeq\,
U^{ac}_{\zt}V_{\bb}t^{c}V^{\dagger}_{\bb}-t^a\,=\,
(U_{\zt} U^\dagger_{\bb})^{ac}t^c-t^a,
\end{align}
since to leading power in $Q_s/P_\perp$, one can approximate  $V_{\bx}\simeq V_{\bb}$ and  $V^{\dagger}_{\yt}\simeq V^{\dagger}_{\bb}$. Unlike in~\cite{Caucal:2024bae},  we do not further assume that the size $R=|\bb-\bz|\sim 1/K_\perp$ of the effective gluon-gluon dipole is much smaller than 
the correlation length $\sim 1/Q_s$ of Wilson lines (this hypothesis permitted the gradient expansion of $U_{\bb}^\dagger$ around $\bz$ in Eq.\,(2.30) in~\cite{Caucal:2024bae}). We therefore stick with the $S$-matrix given by Eq.\,\eqref{Uin}. Using this colour structure, the amplitude for $q\bar q g$ production with the gluon emission occurring before ($B$) the shock-wave reads
\begin{align}\label{PsigBfull}
    \mcal{M}_{B}^{\lambda\bar\lambda\sigma\bar\sigma}&=8ee_fg \sqrt{z_1z_2}q^+\delta^{\sigma,-\bar\sigma}\left(z_2\delta^{\sigma\lambda}-z_1\delta^{\sigma,-\lambda}\right)\et^{\bar\lambda*,n}\et^{\lambda,j}\nn
    &\times \mcal{H}^{mj}(\Pt,z_1,z_2)\mcal{G}_B^{mn}(\Kt,\ellt,\mcal{M}_g) \,, 
\end{align}
where we recall that $\ellt=\Kt+\kgt$ is the transverse momentum transferred by the target, cf. \eqn{defl}, and $\sigma,\bar\sigma,\lambda,\bar\lambda$ respectively refer to the helicity of the quark and the antiquark and to the polarisations of the virtual photon and final state gluon. The (amplitude-level) hard factor describing the momentum distribution of the hard $q\bar q$ pair together with its coupling to the gluon is given by
\begin{align}
	\label{hardamp}
	\mcal{H}^{mj}(\Pt, z_1, z_2) \equiv	\frac{1}{P_\perp^2 + \bar{Q}^2}
	\left(
	\delta^{mj} - 
	\frac{2 \Pt^m \Pt^j}{P_{\perp}^2+\bar{Q}^2}
	\right) \,.
\end{align}
The final ingredient in \eqn{PsigBfull} is the tensor $\mcal{G}_B^{mn}$ which reads
\begin{align}\label{GBfull}
    \mcal{G}_B^{mn}(\Kt,\ellt,\mcal{M}_g)&=\int\frac{\rmd^2\kt}{(2\pi)^2}\,\mathcal{C}^a_B(\kt,\ellt-\kt)\frac{(\kt-\Kt)^m(\kt-\Kt)^n-\frac{1}{2}(\kt-\Kt)^2\delta^{mn}}{(\kt-\Kt)^2+\mcal{M}_g^2} \,, 
\end{align}
where $\mathcal{C}^a_B$ the Fourier transform of the colour structure in Eq.\,\eqref{Uin}:
\begin{align}
    \mathcal{C}^a_B(\ellt,\ellt')&=\int\rmd^2\bb\ \rmd^2\zt \ e^{-i\ellt\cdot\bb-i\ellt'\cdot\bz} \left[ (U_{\zt} U^\dagger_{\bb})^{ac}t^c-t^a\right]\,.
\end{align}
Here, $a$ is the colour index of the final state gluon.
The traceless momentum dependent factor following $ \mathcal{C}^a_B$ in Eq.\,\eqref{GBfull} encodes the transverse momentum distribution of the gluon emitted gluon by the quark-antiquark dipole~\cite{Iancu:2022lcw}. Eq.\,\eqref{PsigBfull} also involves the effective gluon virtuality $\mcal{M}^2_g$, defined as 
\beq\label{Mdef}
	\mcal{M}_g^2\equiv
	z_g\left(Q^2+
	\frac{k_{1\perp}^2}{z_1} + 
	\frac{k_{2\perp}^2}{z_2} 
	\right)=
	{z_g}
	\left(Q^2 +\frac{P_{\perp}^2}{z_1 z_2} 	\right)+\mathcal{O}\left(\frac{K_\perp^2}{P_\perp^2}\right),
	\eeq
which also depends upon the kinematics of the hard $q\bar q $ pair.

\paragraph{Gluon emission after the shock-wave.} Likewise, for gluon emissions  in the final state, the colour structure is 
\begin{align}\label{eq:Uout}
     \left[V_{\bb}\partial^m V_{\bb}^\dagger, t^a\right]=-
\big(U_{\bb}\partial^m U_{\bb}^\dagger\big)^{ac} t^c \,, 
\end{align}
where the commutator structure emerges after combining the gluon emissions by the quark and the antiquark. To obtain the right hand side of Eq.\,\eqref{eq:Uout}, we have also used the identity $V_{\bb}^\dagger t^aV_{\bb}=U_{\bb}^{ac}t^c$. The $q\bar qg$ LCWF for a final state gluon emission then becomes 
\begin{align}\label{PsigAfull}
    \mcal{M}_{A}^{\lambda\bar\lambda\sigma\bar\sigma}&=8iee_fg \sqrt{z_1z_2}q^+\delta^{\sigma,-\bar\sigma}\left(z_2\delta^{\sigma\lambda}-z_1\delta^{\sigma,-\lambda}\right)\et^{\bar\lambda*,n}\et^{\lambda,j}\nn
    &\times \mcal{H}^{mj}(\Pt,z_1,z_2)\mcal{G}_A^{mn}(\Kt,\ellt) \,, 
    \end{align}
where the subscript $A$ stands for ``after'' and (it is understood that $\ellt=\Kt+\kgt$, cf. \eqn{defl})
\begin{align}
	\label{GmjF}
    \mcal{G}^{mn}_A(\Kt,\ellt)
	&\,\equiv \frac{\kgt^n}{k_{g\perp}^2}\,\mathcal{C}^{a,m}_A(\ellt)\,,\quad \mathcal{C}^{a,m}_A(\ellt)=\int\rmd^2\bb \ e^{-i\ellt\cdot\bb}\big(U_{\bb}\partial^m U_{\bb}^\dagger\big)^{ac} t^c \,.
\end{align}

The complete amplitude for $q\bar qg$ production in DIS at small $x$ and leading power in $K_\perp^2/P_\perp^2$ and $Q_s^2/P_\perp^2$ is the sum of Eq.\,\eqref{PsigBfull} and Eq.\,\eqref{PsigAfull}.  A similar expression has been previously obtained in~\cite{Xiao:2017yya} at the amplitude level, working directly at operator level.

\subsection{Real gluon emission at leading power: the cross-section}

We will now demonstrate that, after taking the modulus squared of the full amplitude, summing over the polarisation, spin and colour indices, integrating over the gluon longitudinal momentum ($z_g$) and over its transverse momentum $\kgt$ (or, equivalently, over the transferred momentum $\ellt=\Kt+\kgt$), and performing an appropriate change of the longitudinal integration variable (from $z_g$ to $\xi$, see below),  one finds a TMD-factorised contribution to the dijet cross-section with the same hard factor as at tree-level and a correction $ \Delta \mcal{W}_{\mcal{R}}^{mn}(x, \Kt, P_\perp^2) $ to the WW gluon TMD. As we shall see, this correction depends upon the hard scale $\PT^2$ via the boundaries of the integral over $z_g$ (or $\xi$).

The real contribution to the differential cross-section for inclusive back-to-back $q\bar q$ production is given by\footnote{We \textit{average} over the transverse polarisations $\lambda=\pm1$ of the virtual photon.}
\begin{align}
    \frac{\dif \sigma^{\gamma_{T}^* A \to q\bar{q}(g)X}_{\mcal{R}}}
	{\rmd z_1
  	\rmd z_2\dif^2\Pt\dif^2\Kt} &\,=\frac{1}{16z_1z_2(q^+)^2(2\pi)^8}\delta(1-z_1-z_2)\nn
    &\times\int\frac{\dif z_g}{z_g}\int\dif^2\kgt \left\langle\frac{1}{2}\sum_{\lambda\bar\lambda\sigma\bar\sigma}\left|\mcal{M}_{A}^{\lambda\bar\lambda\sigma\bar\sigma}+\mcal{M}_{B} s^{\lambda\bar\lambda\sigma\bar\sigma}\right|^2\right\rangle_X\,,\label{eq:sigma1R}
\end{align}
where the boundaries for the $z_g$ integration have been anticipated in the introduction and will be shortly explained in more detail.
The external brackets denote the CGC average over the gluon distribution evolved down to $X\equiv x_{q\bar q g}$ with
\beq\label{xgdef}
  x_{q\bar q g} =\frac{1}{2q^+P_N^-}\left(Q^2+\frac{k^2_{1\perp}}{z_1} +\frac{k^2_{2\perp}}{z_2} 
 +\frac{k^2_{g\perp}}{z_g}\right)=\,\frac{Q^2+\frac{P^2_{\perp}}{z_1z_2} 
  +\frac{k^2_{g\perp}}{z_g}}{W^2+Q^2}+\mathcal{O}\left(\frac{K_\perp^2}{P_\perp^2}\right).\eeq
Notice that $X$ is not the same as the argument $x$ of the TMD in the LO formula Eq.\,\eqref{sigma0}: one rather has $x=x_{q\bar q} < X=x_{q\bar q g}$. Yet, they can be identified with each other so long as they both refer to the
BK/JIMWLK evolution of the relevant color correlators. Indeed, when computing the cross-section at NLO, it suffices to consider the leading-order version of high-energy evolution --- that is, the BK/JIMWLK equations to leading logarithmic accuracy in $\ln(1/x)$. This means that the difference between $\ln(1/x_{q\bar q})$ and $\ln(1/x_{q\bar q g})$ can be neglected provided $\alpha_s\ln(x_{q\bar q g}/x_{q\bar q})\ll 1$, a condition that we implicitly assume throughout (see also the discussion in Sect.~\ref{sub:ms-DMMX}).

Clearly, the cross-section in \eqn{eq:sigma1R} involves three types of  contributions: direct initial-state emissions, direct final-state emissions, and interference terms. We will consider these three contributions one after the other.

The contribution from initial-state emissions alone is computed from the modulus squared of $\mcal{M}_{B}$ given by \eqn{PsigBfull} in Eq.\,\eqref{eq:sigma1R}. The result takes a factorised form,
\begin{align}\label{sigT}
    \frac{\dif \sigma^{\gamma_{T}^* A \to q\bar{q}(g)X}_{B}}
	{\rmd z_1
  	\rmd z_2\dif^2\Pt\dif^2\Kt}= & \  \alpha_{em}\alpha_s e_f^2\delta(1-z_1-z_2)\frac{z_1^2+z_2^2}{(P_\perp^2+\bar Q^2)^2}\left[\delta^{mn}-\frac{4\bar Q^2\Pt^m\Pt^n}{(P_\perp^2+\bar Q^2)^2}\right]\nonumber\\
    &\times \Delta \mcal{W}_{B}^{mn}(x, K_{\perp}) \,, 
\end{align}
with the tensorial distribution $\Delta \mcal{W}_{B}^{mn}(x, K_{\perp})$ defined as 
\begin{align}
		\label{DeltaFB0}
\Delta \mcal{W}_{B}^{mn}(x, K_{\perp}) =\frac{1}{2\pi^4} \int\frac{\rmd^2\kgt}{(2\pi)^2}\int \frac{\rmd z_g}{z_g}\,\big\langle \mcal{G}^{mj }_B(\Kt,\ellt,\mcal{M}_g)\,
\mcal{G}^{nj\,* }_B(\Kt,\ellt,\mcal{M}_g)\big\rangle_{X}\,.
\end{align}
As earlier mentioned, to the accuracy of interest one can
ignore the $\kgt$ dependence in the variable $X=x_{q\bar qg}$ when this appears as a rapidity argument for the high-energy evolution. Then, the integral over $\kgt$ (or equivalently $\ellt$) identifies the gluon transverse coordinates, $\bz$ and $\bar{\bz}$, in the DA and the CCA respectively. The colour structure of the cross-section becomes (the trace $\rm tr$ refers to the fundamental representation)
\begin{align}
   \left\langle {\rm tr}\,\Big[\big(U_{\bz} U^\dagger_{\bb}\big)^{ac}t^c-t^a\Big]
    \Big[\big(U_{\bz} U_{\bbb}^\dagger\big)^{ad}t^d-t^a\Big]\right\rangle_{X}&=\frac{N_c^2-1}{2}\int\rmd^2 \qt\,\tilde{\mcal{D}}(X, \Bt, \qt)\nn
    &\hspace{-1cm}\times \left[e^{-i \qt \cdot (\bbb-\bb)}- e^{-i \qt \cdot (\bz-\bb)}
- e^{i \qt \cdot(\bz -\bbb)} +1\right] \,,
\end{align}
where $\tilde{\mcal{D}}(X, \Bt, \kt)$ is the Fourier transform of the colour dipole in the adjoint representation
\begin{align}
    \tilde{\mcal{D}}(X, \Bt, \qt)\equiv \int\frac{\rmd^2\rt}{(2\pi)^2} \ e^{i \qt\cdot\rt} \frac{1}{N_c^2-1}\  \left\langle {\rm Tr}\left[U_{\Bt+\frac{\rt}{2}} U^\dagger_{\Bt-\frac{\rt}{2}} \right]\right\rangle_X \,,
    \label{tildeD}
\end{align}
with $\textrm{Tr}$ the trace in the adjoint representation
and $\Bt$ the dipole impact parameter. Note that the function $\tilde{\mcal{D}}(X, \Bt, \qt)$ is real.
The integrals over $\bb$ and $\bbb$ give two delta functions, which are used to simplify the integrals over $\kt$ and $\overline{\kt}$ in the modulus square of Eq.\,\eqref{GBfull}:
\begin{align}
    \int\rmd^2\bb\rmd^2\bbb \ [...]\rightarrow &(2\pi)^4\left[\delta^{(2)}(\qt-\kt)\delta^{(2)}(\qt-\overline{\kt})-\delta^{(2)}(\qt-\kt)\delta^{(2)}(\overline{\kt}) \right. \nonumber \\
    & \left. -\delta^{(2)}(\kt)\delta^{(2)}(\qt-\overline{\kt})+\delta^{(2)}(\kt)\delta^{(2)}(\overline{\kt})\right]\,.
\end{align}
Hence,
\begin{align}\label{eq:DeltaFb1}
        &\Delta \mcal{W}_{B}^{mn}(x, \Kt) =\frac{C_FN_c}{8\pi^4}\int\frac{\rmd z_g}{z_g}\int\rmd^2\Bt\int\rmd^2\qt\,\tilde{\mcal{D}}(X, \Bt, \qt)\left\{\frac{(\qt-\Kt)^4\delta^{mn}}{((\qt-\Kt)^2+\mcal{M}_g^2)^2}\right.\nn
    &-\left[\frac{2(\qt-\Kt)^m (\qt-\Kt)^j- \delta^{mj}(\qt-\Kt)^2}
	{(\qt-\Kt)^2 +\mcal{M}_g^2}\right]\left[
	\frac{2\Kt^n \Kt^j- \delta^{nj} K_{\perp}^2}
	{K_{\perp}^2 +\mcal{M}_g^2}\right]+\frac{K_\perp^4\delta^{mn}}{(K_\perp^2+\mcal{M}_g^2)^2}\nonumber\\
    &\left.- \left[\frac{2\Kt^m\Kt^j- \delta^{mj}K_\perp^2}
	{K_\perp^2 +\mcal{M}_g^2}\right]\left[
	\frac{2(\qt-\Kt)^n (\qt-\Kt)^j- \delta^{nj} (\qt-\Kt)^2}
	{(\qt-\Kt)^2 +\mcal{M}_g^2}\right]\right\}\,.
\end{align}

At this level, it is useful to observe that the variable $\qt$ introduced via the Fourier transform in \eqn{tildeD} is, be definition, the transverse momentum transferred by the target to the $q\bar q g$ system in the approximations of interest. Hence, this plays the same role as the variable $\ellt$ that we have earlier integrated out. Accordingly, the difference $ \qt-\Kt$ can be identified with the transverse momentum $\kgt=\ellt-\Kt$ carried by the unmeasured gluon in the $s$-channel.

By inspection of \eqn{eq:DeltaFb1}, one can verify that, to leading power in $1/P_\perp$, the integrals over $\qt$ and $z_g$ are controlled by $q_\perp\lesssim \KT$ and, respectively, by relatively low values for $z_g$, for which $\mcal{M}_g^2\lesssim \KT^2$ --- meaning, $z_g\lesssim K_{\perp}^2/P_\perp^2$ (recall \eqn{Mdef}). Indeed, in the opposite situation where $z_g\gg K_{\perp}^2/P_\perp^2$, one can write e.g.
\begin{align}
\frac{K_{\perp}^2}{K_{\perp}^2 +\mcal{M}_g^2}\,\simeq \,\frac{K_{\perp}^2}{\mcal{M}_g^2}\,
\sim \,\frac{1}{z_g} \frac{K_{\perp}^2}{P_\perp^2} \,\ll \,1.\end{align} 
Moreover, it is easy to see that the integral over $z_g$ becomes logarithmic 
in the regime where $z_g$ is {\it much} smaller, such that  $z_g\ll K_{\perp}^2/P_\perp^2$. Indeed, in this case, one can ignore the gluon virtuality $\mcal{M}_g^2$ in the denominators and then the only remaining dependence upon $z_g$ is in the integration measure ${\rmd z_g}/{z_g}$. As we shall demonstrate
in Sect.~\ref{sub:ms-DMMX}, this logarithmic integration contributes to the B-JIMWLK evolution of the WW gluon TMD.

\eqn{sigT} is strongly suggestive of TMD factorisation: it is written as the product between a hard factor (the same as at tree-level, compare with \eqn{HLO}) encoding the kinematics of the $q\bar q$ dijet and a tensorial distribution $\Delta \mcal{W}_{B}^{mn}(x, K_{\perp})$ which depends upon the target variables $x$ and $K_{\perp}$. But the reality is more subtle. First, the factorisation is not yet complete, since the tensor $\Delta \mcal{W}_{B}^{mn}(x, K_{\perp})$ also depends upon the variables $z_i$, $\PT$, and $Q$ of the hard $q\bar q$ pair, via the effective gluon virtuality $\mcal{M}^2_g$, as shown in \eqn{Mdef}. Second, there is an additional dependence upon $P_\perp$, which enters via the integration limits on $z_g$, to be discussed in the next section. Here, we would like to show that the breaking of factorisation by the gluon virtuality $\mcal{M}^2_g$ is  an artifact, that can be easily remediated via a change of variables. This change of variables is also needed for the physical interpretation of $\Delta \mcal{W}_{B}^{mn}(x, K_{\perp})$ as an evolution of the {\it target}.

Specifically, let us replace $z_g$ (the longitudinal momentum fraction of the gluon  w.r.t. the photon) with the variable $\xi$, which is defined as 
\beq\label{Mxi2}
\xi\equiv  \frac{x_{q\bar q }}{x_{q\bar q g}}\equiv\frac{x}{X}\,.\eeq
and represents a longitudinal momentum fraction w.r.t. the target.
Using the definitions of $x_{q\bar q }$ and $x_{q\bar qg}$ in Eqs.\,\eqref{xqqdef} and \eqref{xgdef}, one finds
\begin{align}
    z_g 	=\frac{\xi}{1-\xi} \frac{\kg^2}{Q^2+{P_\perp^2}/{(z_1z_2)}}
\quad \Rightarrow \quad
\mcal{M}_g^2 = \frac{\xi}{1-\xi} \kg^2\,.
\label{eq:zg-to-xi}
\end{align}
After this change of variable (which is possible because we integrate out the gluon), the effective virtuality $\mcal{M}_g^2$ becomes a function of the target kinematical variables $\xi$ and $\Kt$.

It is furthermore convenient to also change the transverse integration variable, from  $\ellt$ to $\kgt\equiv \qt-\Kt$, and to introduce the leading-order dipole gluon TMD, defined as~\cite{Dominguez:2011wm} 
\begin{align}
   \mcal{F}_{g}^{D,(0)}(x,\ellt)\equiv \frac{C_F}{2\pi^2\alpha_s}\ell_\perp^2\int\rmd^2\Bt \   \tilde{\mcal{D}}(x, \Bt, \ellt) \,.
   \label{eq:gluon-dipole}
\end{align}

Putting these results together and using $\rmd z_g/z_g = \rmd \xi/[\xi(1-\xi)]$ for the change in the longitudinal variable, one finds the ``initial-state'' contribution in the form
\begin{align}
	\label{DeltaFB}
	\Delta \mcal{W}_{B}^{mn}(x, \Kt)&=\frac{\alpha_s N_c}{\pi^2}
\int\frac{\rmd^2 \kgt}{{(\Kt+\kgt)^2}}\int\rmd\xi\,\frac{1-\xi}{4\xi}\ 
 \mcal{F}_{g}^{D,(0)}\left(\frac{x}{\xi},  \Kt+\kgt\right)\nn*[.2cm]
&\,\times\left\{\delta^{mn}\left[1+\frac{\KT^4}{[(1-\xi)\KT^2+\xi\kg^2]^2}\right]-
\frac{2\KT^2}{(1-\xi)\KT^2+\xi\kg^2}\right.\nn
&\left.\times \bigg[\delta^{mn}
- \frac{2\kgt^m \kgt^n}{\kg^2} -\frac{2\Kt^m \Kt^n}{\KT^2}+ \frac{2(\kgt^m\Kt^n+\kgt^n\Kt^m) (\kgt\cdot\Kt)}{\kg^2\KT^2}\bigg]\right\} \,,
\end{align}
where the integration limits on $\xi$ will be specified in the next section. In general, \eqn{DeltaFB} involves a non-trivial convolution in $\xi$  between the $\xi$-dependent kernel inside the curly brackets and the dipole gluon TMD whose high-energy evolution must be evaluated at the longitudinal scale $X=x/\xi$ (see Eq.\,\eqref{Mxi2}). That said, within the approximations of interest, one can replace $X\simeq x$ (i.e. $x_{q\bar q }\simeq x_{q\bar qg}$) within $ \mcal{F}_{g}^{D,(0)}$ and then the $\xi$-integral runs over the kernel alone.

Consider similarly the contribution from final-state emissions alone, i.e.~$|\mcal{M}_A|^2$, which is much simpler: the analog of \eqn{DeltaFB} reads
\begin{align}
	\label{DeltaFA}
\Delta \mcal{W}_{A}^{mn}(x, \Kt)&
 =\frac{1}{2\pi^4} \int\frac{\rmd^2\kgt}{(2\pi)^2}\int \frac{\rmd z_g}{z_g}\,\big\langle \mcal{G}^{mj }_A(\Kt,\ellt)\,
\mcal{G}^{nj\,* }_A(\Kt,\ellt)\big\rangle_{X}\nn 
&=\frac{1}{4\pi^4} \int\frac{\rmd^2\kgt}{(2\pi)^2}\frac{1}{\kgt^2}\int \frac{\rmd z_g}{z_g}\int\rmd^2\bb\rmd^2\bbb e^{-i\ellt\cdot(\bb-\bbb)}\left\langle \big(U_{\bb}\partial^m U_{\bb}^\dagger\big)^{ac}\big(\partial^nU_{\bbb} U^\dagger_{\bbb} \big)^{ca}\right\rangle_X\nonumber\\
&=\frac{\alpha_sN_c}{\pi^2}\int\frac{\rmd^2\kgt}{\kgt^2} \int\frac{\rmd \xi}{\xi(1-\xi)}\mcal{W}^{mn,(0)}\left(\frac{x}{\xi},\Kt+\kgt\right) \,.
\end{align}
Note that for this contribution, $\kgt$ is truly the transverse momentum of the gluon measured in the final state. Moreover the integral over $z_g$ is logarithmic over the whole region in phase-space to which \eqn{DeltaFA} applies, which extends up to $z_g\sim \kg/\PT$ (see the discussion in the next section). After the change of variables from $z_g$ to $\xi$, this generates logarithmic domains of integration near both endpoints, $\xi\ll 1$ and $1-\xi\ll 1$, which in turn will be seen to contribute to the B-JIMWLK and the CSS evolutions, respectively.

Finally, the contribution from interferences between initial and final state gluon emissions is given by $\propto \mcal{M}_B\mcal{M}_A^*+\mcal{M}_A\mcal{M}_B^*$. This term involves a colour operator which is neither the gluon dipole nor the Weiszäcker-Williams operator:
\begin{align}
    \left\langle \textrm{tr} \left\{\left[ (U_{\zt} U^\dagger_{\bb})^{ac}t^c-t^a\right]\left[(\partial^n U_{\bbb})U_{\bbb}^\dagger\right]^{da} t^d \right\}\right\rangle_X=\frac{1}{2}\left\langle \textrm{Tr} \left[ U_{\bz}U^\dagger_{\bb}(\partial^nU_{\bbb})U^\dagger_{\bbb} \right]  \right\rangle_X \,.
\end{align}
We thus introduce the double Fourier transform of this new operator, which carries the transverse index $n=1,2$:
\begin{align}
      \mcal{Z}^{n,(0)}(X,\qtone,\qttwo)&\equiv\frac{C_F\qttwo^2}{2\pi^2\alpha_s}\int\rmd^2\Bt\int\frac{\rmd^2\rtone}{(2\pi)^2}\frac{\rmd^2\rttwo}{(2\pi)^2}e^{i\qtone\cdot\rtone+i\qttwo\cdot\rttwo}\nn
      &  \times \frac{i}{N_c^2-1}\left\langle \textrm{Tr} \left[ U_{\Bt+\rttwo/2}U^\dagger_{\Bt-\rttwo/2}\partial^n U_{\Bt+\rttwo/2+\rtone}U^\dagger_{\Bt+\rttwo/2+\rtone} \right] \right\rangle_X\,.\label{eq:Zop-def}
\end{align}
The overall prefactor has been chosen so that we have the simple identity relating $\mathcal{Z}^n$ to the dipole gluon TMD:
\begin{align}
    \int\rmd^2\qtone \mcal{Z}^{n,(0)}(x,\qtone,\qttwo)= q_2^n\,\mcal{F}_g^{D,(0)}(x,\qttwo)\,,\label{eq:Zop-id}
\end{align}
which will be useful later. With this definition, the interference term reads
\begin{align}
	\label{DeltaFint}
	&\Delta \mcal{W}_{\rm int}^{mn}(x, \Kt)=\frac{\alpha_s N_c}{2\pi^2}
\int\frac{\rmd^2 \kgt\rmd^2\kt}{\kt^2}\int\frac{\rmd\xi}{\xi}\ 
 \mcal{Z}^{n,(0)}\left(\frac{x}{\xi},\Kt+\kgt,\kt\right)\nn*[.2cm]
&\,\times\frac{\kgt^i(\Kt-\kt)^2}{\kgt^2\left[(1-\xi)(\Kt-\kt)^2+\xi\kgt^2\right]}\left[\delta^{mi}-\frac{2(\Kt-\kt)^m(\Kt-\kt)^i}{(\Kt-\kt)^2}\right] +\overline{(m\leftrightarrow n)}\,.
\end{align}
As in the final state contribution, the integration variable $\kgt$ in Eq.\,\eqref{DeltaFint} is also the transverse momentum of the radiated gluon. The term $\overline{(m \leftrightarrow n)}$ is obtained from the first term by swapping the indices $m$ and $n$, and then taking the complex conjugate.

Including the interferences between the initial state and final state gluon emissions, our final result for the real contribution to the inclusive $q\bar q$ dijet cross-section in DIS in the back-to-back limit reads
\begin{align}
 \Delta \mcal{W}_{\mcal{R}}^{mn}(x, \Kt)&=\Delta \mcal{W}_{A}^{mn}(x, \Kt)+\Delta \mcal{W}_{B}^{mn}(x, \Kt)+\Delta \mcal{W}_{\rm int}^{mn}(x, \Kt)\,.
\end{align}
This yields the following correction to the
inclusive back-to-back $q\bar q$ pair production in DIS:
\begin{align}\label{Dsigma}
    \frac{\dif \sigma^{\gamma_{T}^* A \to q\bar{q}(g)X}_{\mcal{R}}}
	{\rmd z_1
  	\rmd z_2\dif^2\Pt\dif^2\Kt}= & \  \alpha_{em}\alpha_s e_f^2\delta(1-z_1-z_2)\frac{z_1^2+z_2^2}{(P_\perp^2+\bar Q^2)^2}\left[\delta^{mn}-\frac{4\bar Q^2\Pt^m\Pt^n}{(\Pt^2+\bar Q^2)^2}\right]  \Delta \mcal{W}_{\mcal{R}}^{mn}(x, \Kt)\,.
\end{align}
This result shows that the NLO contribution we have identified in this paper can be written in a TMD factorised form, with the same hard factor as at LO; yet, the intervening TMD distributions involve additional colour operators besides the WW gluon TMD. This was to be expected from our experience with the high-energy evolution of the WW correlator, to be derived from our formalism in section~\ref{sub:ms-DMMX}.

\subsection{Phase-space constraints on the longitudinal fraction $\xi$}
\label{sec:phase-space-constraints}

The boundaries of the integral over $\xi$ are inherited from those over $z_g$, via the change of variables \eqref{eq:zg-to-xi}. Besides ensuring that $\xi$ (or $z_g$) spans the appropriate region in phase-space for the dijet problem at hand, these boundaries also provide physical cutoffs for the end-point divergences at $\xi=0$ and $\xi=1$ which are a priori present in the various integrals contributing to
$\Delta \mcal{W}_{\mcal{R}}^{mn}(x, K_{\perp}, P_\perp^2)$ (see e.g. the final-state contribution in \eqn{DeltaFA}).  First, $x/\xi$ should be smaller than $x_0$, which is the largest value of $x$ for which the CGC approach can be still applied --- a common choice is $x_0\sim 10^{-2}$. (In particular, the initial conditions for the BK or B-JIMWLK equations are formulated at $x_0$.) This immediately implies $\xi>x/x_0$, where it is understood that $x\le x_0$. Second, the real gluon should propagate outside of the two jets made by the quark and the antiquark. As we show now, this last condition regulates the $\xi\to 1$ divergence in $\Delta \mcal{W}_{\mcal{R}}^{mn}(x, \Kt)$ which is generated by final-state gluon emissions alone. Using jet definitions from the generalised $k_t$ family~\cite{Ellis:1993tq,Dokshitzer:1997in,Wobisch:1998wt,Cacciari:2008gp,Cacciari:2011ma}, as standard in hadronic collisions, the condition for the gluon to remain outside the jet initiated by its (quark or antiquark) emitter reads (in the small angle limit),
    \begin{align}
     \left(\kgt-\frac{z_g}{z_1}\Pt\right)^2\simeq \kgt^2\ge R^2z_g^2P_\perp^2   \,.
    \end{align}
where $R < 1$ is the jet radius parameter and we have used the fact that, say, the quark jet is a cone with opening angle $\Delta \theta_q \simeq R(k_{1\perp}/z_1q^+) \sim RP_\perp/q^+$ centred along the quark propagation angle $\theta_q \sim  P_\perp/q^+$. (We have also used the fact that $k_{1\perp}\simeq P_\perp$ and $z_1\sim \order{1/2}$ for back-to-back dijets.) Then the final inequality in the equation above implies $z_g < \kg/(RP_\perp)\sim \kg/P_\perp$ (for $R$ not {\it too} small), which in turn justifies the approximation we made within the original parenthesis.

 In term of the target variable $\xi$ (see Eq.\,\eqref{eq:zg-to-xi}), this condition becomes
    \begin{align}\label{xi0}
        \xi\le 1-\xi_0(\kg),\,\quad \xi_0(\kg)\equiv\frac{\kg P_\perp R}{Q^2+M_{q\bar q}^2} \,.
    \end{align}
Physically, $\xi_0(\kg)\sim \kg/P_\perp\ll 1$ is a lower cutoff on the longitudinal fraction $1-\xi$ of the unmeasured gluon emitted in the $s$-channel. This cutoff depends upon the gluon transverse momentum $\kg$ and also upon the hard scales $P_\perp$ and $Q$, although this last dependence is left implicit in our notations.

We thus see that, via the rapidity cutoff $\xi_0$, the correction $\Delta \mcal{W}_{\mcal{R}}^{mn}$ to the WW gluon TMD acquires a dependence upon the hard scales. Our final result for this  correction reads
\begin{align}\label{eq:DF1R-full}
     &\Delta \mcal{W}_{\mcal{R}}^{mn}(x, \Kt,P_\perp^2)=\frac{\alpha_sN_c}{\pi^2}\int\rmd^2\kgt\int_{x/x_0}^{1-\xi_0(\kg)}\rmd \xi\Bigg\{\frac{1}{\xi(1-\xi)}\frac{1}{\kgt^2}\mcal{W}^{mn,(0)}\left(\frac{x}{\xi},\Kt+\kgt\right)\nn
& +\frac{1-\xi}{4\xi}\frac{1}{(\Kt+\kgt)^2}\left[\frac{\xi^2(\KT^2-\kg^2)^2\delta^{mn}}{\left((1-\xi)\KT^2+\xi\kg^2\right)^2}\right.\nonumber\\
&\left.+
\frac{4\KT^2}{(1-\xi)\KT^2+\xi\kg^2}\bigg(\frac{\kgt^m \kgt^n}{\kg^2} +\frac{\Kt^m \Kt^n}{\KT^2}- \frac{(\kgt^m\Kt^n+\kgt^n\Kt^m) (\kgt\cdot\Kt)}{\kg^2\KT^2}\bigg)\right]\mcal{F}_{g}^{D,(0)}\left(\frac{x}{\xi},  \Kt+\kgt\right)\nn
 &+\frac{1}{2}\Bigg[\int\frac{\rmd^2\kt}{\kt^2}\frac{1}{\xi}\mcal{Z}^{n,(0)}\left(\frac{x}{\xi},\Kt+\kgt,\kt\right)  \nn
&\times \frac{\kgt^i(\Kt-\kt)^2}{\kg^2\left[(1-\xi)(\Kt-\kt)^2+\xi\kg^2\right]}\left[\delta^{mi}-\frac{2(\Kt-\kt)^m(\Kt-\kt)^i}{(\Kt-\kt)^2}\right]+\overline{(m\leftrightarrow n)} \Bigg]\Bigg\} \,.
\end{align}
This expression represents the main result of this paper. It describes a gluon splitting $g\to gg$ occurring in the target wavefunction in which  the parent  gluon with transverse momentum $\ellt$ and target longitudinal momentum fraction $x/\xi$ decays into into two gluons with transverse momenta $\Kt$, $\kgt=\ellt-\Kt $ and longitudinal momentum fractions $\xi,1-\xi$. It holds for generic values $\KT,\kg \ll \PT$ and it includes saturation effects. 


The TMD splitting of an unpolarised WW gluon can be obtained from Eq.\,\eqref{eq:DF1R-full} by contracting the equation with $\delta^{mn}$. Defining
\begin{align}
    \Delta \mcal{F}_{g,\mcal{R}}^{WW}\equiv \Delta \mcal{W}_{\mcal{R}}^{mn}\delta^{mn} \,,
\end{align}
we have 
\begin{align}\label{eq:DF1R-full-unpol}
     &\Delta \mcal{F}_{g,\mcal{R}}^{WW}(x, \Kt,P_\perp^2)=\frac{\alpha_sN_c}{\pi^2}\int\rmd^2\kgt\int_{x/x_0}^{1-\xi_0(\kg)}\rmd \xi\Bigg\{\frac{1}{\xi(1-\xi)\kgt^2}\mcal{F}_g^{WW,(0)}\left(\frac{x}{\xi},\ellt\right)\nn
&+\frac{1-\xi}{2\xi\ell_\perp^2}\left[\frac{\xi^2(\KT^2-\kg^2)^2}{\left((1-\xi)\KT^2+\xi\kg^2\right)^2}+
\frac{4\KT^2}{(1-\xi)\KT^2+\xi\kg^2}\bigg(1- \frac{(\kgt\cdot\Kt)^2}{\kg^2\KT^2}\bigg)\right]\mcal{F}_{g}^{D,(0)}\left(\frac{x}{\xi},  \ellt\right)\nn
&+\int \frac{\rmd^2\kt}{\kt^2} \frac{1}{\xi} \frac{(\Kt-\kt)^2\ \kgt^i \left[\delta^{mi}-\frac{2 (\Kt-\kt)^{m} (\Kt-\kt)^{i}}{(\Kt-\kt)^2}\right]}{\kg^2\left((1-\xi)(\Kt-\kt)^2+\xi\kg^2\right)}\mcal{Z}^{m,(0)}\left(\frac{x}{\xi},\ellt,\kt\right) \Bigg\} \,,
\end{align}
with as before, $\ellt=\Kt+\kgt$. A similar expression can obtained for linearly polarised gluons; this is discussed in detail in Appendix~\ref{app:linpol-TMD}.

\section{Unveiling the QCD evolutions equations}
\label{sec4}

In this section, we aim at simplifying Eq.\,\eqref{eq:DF1R-full} in three kinematical regimes: \texttt{(i)} the small-$x$ limit $x\ll \xi \ll 1$, without any assumption on the relative magnitude between $K_\perp,\ell_\perp$ and $Q_s$, \texttt{(ii)} the DGLAP regime, with strong ordering in transverse momenta, i.e.~$K_\perp\gg \ell_\perp$, but no constraint on $\xi$, \texttt{(iii)} the dilute limit, where $Q_s$ is assumed to be much smaller than both $K_\perp$ and $\ell_\perp$, with $K_\perp\sim \ell_\perp$ being the regime of interest in that case. In these three regimes, we provide physical interpretation of the results (see also Fig.\,\ref{fig:s-vs-t-channel-picture-nlo}) and discuss their connection with the state of the art in the literature.

\subsection{The B-JIMWLK evolution of the Weiszäcker-Williams gluon TMD}
\label{sub:ms-DMMX}

We first investigate the high energy, or small $x$, limit of Eq.\,\eqref{eq:DF1R-full}. When $x/x_0\ll 1$, the integral over $\xi$ develops a logarithmic singularity at the lower limit and to leading logarithmic accuracy is controlled by small values $\xi\ll 1$. Then we can take $\xi=0$ everywhere, except in the arguments of the gluon correlators (which should be kept at $x/\xi$). In the end, the $\xi$ dependence in the kernel becomes simply $1/\xi$. Besides, we can ignore the rapidity regulator $\xi_0(\kg)$ in the upper limit. The expression thus obtained can be recognised as the integral version of an evolution equation of the B-JIMWLK type. To better identify this equation, it is preferable to write its differential version. To that aim, we need to view the gluon emission as just one step in the evolution with decreasing $x$. For simplicity, we take this to be the first step away from the initial condition formulated at $x_0$. In practice this means that $x$ must be smaller, but not {\it much} smaller, than $x_0$, namely such that $\alpha_s\ln(x_0/x)\ll 1$. In this case, one can also ignore the $\xi$-dependence of the color correlators: to leading log accuracy, one can ignore their evolution between $x_0$ and $x$, meaning that one can approximate their rapidity argument as $x/\xi\simeq x \simeq x_0$.

We thus get, after also changing the integration variable $\kgt$ to $\ellt=\kgt+\Kt$,
\begin{align}
     &\Delta \mcal{W}_{\mcal{R}}^{mn}(x, \Kt,P_\perp^2)\underset{x\ll1}{\,\simeq\,}\frac{\alpha_sN_c}{\pi^2}\ln\left(\frac{x_0}{x}\right)\int\rmd^2\ellt\left\{\frac{\mcal{W}^{mn,(0)}\left(x,\ellt\right)}{(\ellt-\Kt)^2}+\frac{1}{\ell_\perp^2}\mcal{F}_{g}^{D,(0)}\left(x,  \ellt\right)\right.\nn*[0.2cm]
     &\times\left[
\frac{(\ellt-\Kt)^m (\ellt-\Kt)^n}{(\ellt-\Kt)^2}+\frac{\Kt^m \Kt^n}{\KT^2}- \frac{((\ellt-\Kt)^m\Kt^n+(\ellt-\Kt)^n\Kt^m) ((\ellt-\Kt)\cdot\Kt)}{(\ellt-\Kt)^2\KT^2}\right]\nn*[0.2cm]
&- \Bigg[\int\frac{\rmd^2\kt}{k_\perp^2}\mcal{Z}^{n,(0)}\left(x,\ellt,\kt\right)\frac{(\ellt-\Kt)^i}{(\ellt-\Kt)^2}\left.\frac{(\Kt-\kt)^m(\Kt-\kt)^i}{(\Kt-\kt)^2}+\overline{(m\leftrightarrow n)} \Bigg]\right\},\label{eq:rDMMX}
\end{align}
where in the interference contribution we used the identity
\begin{align}
    \int \frac{\der^2 \kt}{\kt^2} \mcal{Z}^{n,(0)}\left(x,\ellt,\kt\right) = 0 \,.
\end{align}
Although not obvious at first sight, this is exactly the real part of one step in the JIMWLK evolution of the Weiszäcker-Williams TMD, as described by the DMMX equation~\cite{Dominguez:2011gc}. This is explicitly demonstrated in the appendix~\ref{app:DMMX}. Unlike in~\cite{Dominguez:2011gc} though, this contribution is here written in transverse momentum space. It is also manifestly a non-closed equation, since the right hand side involves not only the WW gluon TMD, but also the gluon dipole TMD and the 3-point correlator $\mcal{Z}^n$. So, this equation is effectively non-linear, as generally the case with the equations from the  Balitsky-JIMWLK hierarchy. Its linearised version, which neglects saturation effects, will be discussed in Sect.~\ref{sec:BFKL}. The full DMMX equation and its (approximate) solutions based on the Gaussian approximation~\cite{Jalilian-Marian:2004vhw,Dominguez:2011wm,Iancu:2011ns,Iancu:2011nj,Lappi:2012nh}
are thoroughly presented in the literature, so we will not pursue here a more detailed analysis.

 Note that thanks to the change of variable from projectile to target longitudinal momentum fraction, the DMMX evolution is naturally formulated in terms of the {\it target} evolution variable $\ln(x_0/x)$. When expressed  in terms of the original longitudinal variable $z_g$, the evolution described by \eqn{eq:rDMMX} corresponds to the so-called {\it kinematically improved} version of the  B-JIMWLK evolution, which is non-local in (projectile) rapidity and constrained by time ordering: successive gluon emissions are strongly ordered both in $z_g$ and in their formation times $\Delta x^+_g=2z_gq^+/\kg^2$~\cite{Beuf:2014uia,Iancu:2015vea,Hatta:2016ujq,Ducloue:2019ezk}. It is interesting to see how this constraint arises in the present context: the above condition $x/x_0\ll \xi \ll 1$ on $\xi$ together with \eqn{eq:zg-to-xi} relating $\xi$ to $z_g$ can be combined to yield the following conditions on $z_g$, or, equivalently, on the gluon formation time (with $s\equiv 2q^+P_N^-$):
 \begin{align}
 \frac{1}{x_0}\,\frac{\kg^2}{s}\,\ll z_g \ll\,\frac{\kg^2}{Q^2}\,,\qquad\mbox{or}
\qquad \frac{1}{x_0 P_N^-}\,\ll \Delta x^+_g\ll\,\frac{2q^+}{Q^2} =\Delta x^+_{\gamma^*}\,,
\end{align}
where we have used $P_\perp^2\sim Q^2$ and we recognised the coherence time $\Delta x^+_{\gamma^*}$ of the virtual photon. These conditions show that the gluon formation time must be much larger than the longitudinal extent $\propto 1/P_N^-$ of the target, but much smaller than the typical lifetime of a $q\bar q$ fluctuation of the virtual photon.
 
 This time-ordering constraint is essential for properly counting  the longitudinal phase-space allocated to the high-energy evolution. It has important consequences in practice: it solves the instability problem which appears in the BK equation at NLO~\cite{Lappi:2015fma,Lappi:2016fmu}
 and it allows to properly isolate the Sudakov logarithms from the other large logarithms present in the NLO corrections to dijet production~\cite{Taels:2022tza,Caucal:2022ulg,Caucal:2023fsf}. Ultimately, it shows that the quantum evolution for dilute-dense collisions is most naturally expressed as the evolution of the dense target wavefunction, even though in practice it is more conveniently computed from the evolution of the dilute projectile. This conclusion applies not only to the high-energy,  B-JIMWLK, evolution that we just described, but also to the collinear, CSS and DGLAP, evolutions to be discussed in what follows.

\subsection{The CSS evolution of the gluon TMD}
\label{sec:CSS}

In this section we investigate the behaviour of the splitting function \eqn{eq:DF1R-full} near $\xi\to 1$ and thus identify a logarithmic divergence that can be recognised as the beginning of the Collins-Soper-Sterman (CSS) evolution~\cite{Collins:1981uk,Collins:1981uw,Collins:1984kg,Collins:2011zzd}. In this context, the  CSS equation describes the evolution of the WW gluon TMD with increasing the hard resolution scale $P_\perp^2$.

The divergence shows up in the limit of a vanishing 
rapidity regulator  $\xi_0\to 0$.  From \eqn{xi0}, we recall that
 \begin{align}\label{xi0Q}
     \xi_0(\kg)\,=\,\frac{\kg}{Q_\perp}\qquad\text{where}\qquad Q_\perp\equiv\,
     \frac{Q^2+M_{q\bar q}^2}{RP_\perp}\,,
    \end{align}
which is indeed small, $\xi_0\ll 1$, in the interesting regime at $Q_\perp\sim P_\perp\gg \kg$.
By inspection of \eqn{eq:DF1R-full}, it is clear that the only would-be singular term when $\xi_0\to 0$ is the first term in the r.h.s., which in the dipole picture corresponds to final-state gluon emissions. We shall shortly isolate the piece proportional to $\ln(1/\xi_0)$, but before doing that it is useful to clarify its physical interpretation, in both the projectile picture and the target picture.

In the dipole picture, the  logarithmic enhancement $\ln(1/\xi_0)$ is generated by the integral over $z_g$ within the interval $\kg^2/P_\perp^2 \ll z_g \ll  k_g/P_\perp$, where we recall that the lower limit represents the boundary with the high-energy evolution, while the upper limit is the condition for the gluon to be emitted outside the jet cone. Hence,  this enhancement can be attributed to gluon emissions at relatively small angles, just outside the measured jets.

In the target picture, the condition on $\xi$ can be written as $1- \xi\ge \xi_0(\kg)$, with  $1- \xi$  the splitting fraction of the gluon emitted in the $s$-channel,  which is not measured. Hence, the logarithm  $\ln(1/\xi_0)$ is generated by very soft gluon emissions which do not significantly alter the longitudinal (minus) momentum of the gluon measured in the $t$-channel.

The would-be singular contribution to \eqn{eq:DF1R-full} in the limit $\xi_0\to 0$  (i.e.~$\xi\to 1$) is easily found as
\begin{align}\label{CSS0}
     \Delta \mcal{W}_{\mcal{R}}^{mn}(x, \Kt, P_\perp^2)\underset{\xi_0\ll1}{=}\,\frac{\alpha_sN_c}{\pi^2}\int\frac{\rmd^2\kgt}
   {\kgt^2}\,\ln\left(\frac{1}{\xi_0}\right)\,\mcal{W}^{mn,(0)}\left({x},\Kt+\kgt\right).
 \end{align}
This can be promoted into an evolution equation by taking a derivative w.r.t. the logarithm $\ln Q_\perp$ of the hard scale. As already mentioned, we assume 
$P_\perp^2\gtrsim Q^2$, hence we can identify $P_\perp$ and $Q_\perp$ and thus write
\beq
\ln\frac{1}{\xi_0} \,=\,\frac{1}{2}\,\ln\frac{P_\perp^2}{\kg^2}\,.\eeq
Then, \eqn{CSS0} can be recognised as the real contribution to the first iteration of the CSS equation.  To obtain the complete equation, one also needs the respective virtual contribution. In the small--$\xi_0$ limit of interest, this is the same as the Sudakov double logarithm computed in~\cite{Taels:2022tza,Caucal:2022ulg,Caucal:2023fsf}. After combing real and virtual contributions, the CSS equation can be written as (for the unpolarised TMD, for definiteness --- the linearly polarised case is presented in appendix~\ref{app:linpol-TMD}):
\begin{align}
	\label{CSSKT}
 \frac{\del \mcal{F}_g(x, \Kt, P_\perp^2)}{\del \ln P_\perp^2}&\,=\frac{\alpha_sN_c}{2\pi}\int\frac{\rmd^2\kgt}{\pi\kgt^2}\, 
  \Big[\mcal{F}_g(x, \Kt+\kgt, P_\perp^2) - \Theta(P_\perp^2-\kg^2)\mcal{F}_g(x, \Kt, P_\perp^2)\Big]. \end{align}
Notice the cancellation of the infrared singularities at $\kg\to 0$ between the real term and the virtual term.

\eqn{CSSKT} is linear, meaning that no saturation effects are included in its structure.  This is in line with the fact that its derivation involved only gluon emissions in the final state, after the collision with the dense target. Yet, saturation effects may still enter via the initial (or boundary) condition and they are expected to be important when $K_\perp\lesssim Q_s$. As argued in~\cite{Caucal:2024bae}, \eqn{CSSKT} should be solved as a boundary value problem,  with the boundary condition formulated at $P_\perp^2\sim \KT^2$ (where the effects of the CSS resummation are negligible). In turn this boundary condition is taken from the CGC effective theory and must include the  B-JIMWLK evolution at small $x$ together with saturation effects for 
 $K_\perp\lesssim Q_s$. This illustrates the fact that the various evolutions that we are about to discuss and which have a common origin --- the gluon splitting vertex in  Eq.\,\eqref{eq:DF1R-full} --- are in fact correlated with each other and must be properly matched.
 
\eqn{CSSKT} is written in transverse momentum space, yet this is not the version of the CSS equation that is commonly shown in the literature (see however~\cite{Shi:2023ejp,vanHameren:2025hyo}). In general one rather prefers to work with its version in transverse {\it coordinate} space, as obtained after a Fourier transform (FT) from $\Kt$ to $\rt$:
\beq\label{FTTMD}
\tilde{\mcal{F}}_g(x,\rt, P_\perp^2)
\,\equiv\int \frac{\rmd^2\Kt}{(2\pi)^2}\,\rme^{i\Kt\cdot\rt}\, \mcal{F}_g(x, \Kt, \PT^2)\,.
\eeq
Clearly, the use of the transverse coordinate representation becomes convenient at high energies, where it facilitates e.g. the resummation of multiple scattering in the eikonal approximation (recall e.g. \eqn{dGWW}). But it is also useful in the more standard context of TMD factorisation at moderate values of $x$, since the respective version of the CSS equation is easier to deal with in practice. One can easily check that the FT of \eqn{CSSKT} takes the simple form (with $c_0\equiv 2\,e^{-\gamma_E}\simeq 1.123$)
\begin{align}
	\label{CSSbT}
 \frac{\del \tilde{\mcal{F}}_g(x, \rt, P_\perp^2)}{\del \ln P_\perp^2}&\,=\,-\frac{\alpha_sN_c}{2\pi}\left[\ln\frac{r_\perp^2P_\perp^2}{c_0}\right]
 \, \tilde{\mcal{F}}_g(x, \rt, P_\perp^2)\,, \end{align}
in which the real piece if formally missing --- in reality, it is responsible for the lower limit $c_0/r_\perp^2$ in the argument of the logarithm. In particular, \eqn{CSSbT} is local in $r_\perp$ and  can be easily solved for a generic boundary condition at $P_\perp^2=
c_0/r_\perp^2$. The effect of the solution is to exponentiate the double Sudakov logarithms in transverse coordinate space:
\begin{align}
	\label{CSSbTsol}
 \tilde{\mcal{F}}_g(x, \rt, P_\perp^2) \,=\, \tilde{\mcal{F}}_0(x, \rt)\,\exp\left\{-\frac{\alpha_sN_c}{4\pi}\,
 \left[\ln\frac{r^2P_\perp^2}{c_0}\right]^2\right\}\,,
 \end{align}
 with $\tilde{\mcal{F}}_0(x, \rt)$ the initial condition of the CSS evolution.
Yet, complications may arise when computing the inverse FT from $\rt$ to $\Kt$, as eventually needed for the cross-sections (see e.g. \eqn{Dsigma}). The respective integral runs over all values of $r_\perp$, including very large values where perturbation theory fails to apply. It is at this level that saturation effects (as included in the boundary condition) could be very helpful: they effectively limit the integral to values $r_\perp\lesssim 1/Q_s$ (see e.g.~\cite{Zheng:2014vka}).

 \subsection{The DGLAP evolution of the gluon PDF}
 
 The DGLAP evolution of the gluon PDF is recovered in the regime where the imbalance of the $q\bar q$ pair is controlled by the recoil from the hard gluon emission: $K_{\perp}\simeq \kg \gg |\Kt+\kgt|$~\cite{Caucal:2024bae}. This in particular requires the dijet imbalance to be quite hard, $K_\perp\gg Q_s$,  and hence less sensitive to gluon saturation. Clearly, this particular region in the transverse phase-space also contributes to the two types of evolutions previously discussed: the B-JIMWLK equation, cf. Sect.~\ref{sub:ms-DMMX}, and the CSS equation, cf. Sect.~\ref{sec:CSS}. Yet, as we shall see, these various evolutions can be unambiguously disentangled in longitudinal phase-space. We have previously seen that the high-energy evolution corresponds to very small values $z_g\ll \kg^2/\PT^2$ for the gluon longitudinal momentum fraction w.r.t. the photon. In this section, we shall consider the complimentary region at $z_g\gtrsim \kg^2/\PT^2$ and we shall compute its contribution in the transverse phase-space where $K_{\perp}\simeq \kg \gg |\Kt+\kgt|$. This will generate  one DGLAP step and one CSS step embedded with each other, with the CSS step evaluated in a leading logarithmic approximation (in the sense of transverse logs). The genuine DGLAP contribution will be then extracted and its precise separation from the B-JIMWLK evolution in longitudinal phase-space will be thoroughly discussed.

 When $K_{\perp}$ and $\kg$ are both large and comparable with each other, it becomes convenient to use the total transverse momentum transferred by the collision, $\ellt=\Kt+\kgt$, as an integration variable instead of $\kgt$ in Eq.\,\eqref{eq:DF1R-full}. We have indeed $K_{\perp}\simeq \kg \gg \ell_\perp$.
 Then most of the terms visible in the integrand of \eqn{eq:DF1R-full} can be simplified by replacing $\kgt$ with $-\Kt$. For the sake of simplicity, we only consider the trace part of $\Delta \mcal{W}_{\mcal{R}}^{mn}(x, \Kt ,P_\perp^2)$, that is, the real NLO correction to the unpolarised WW gluon TMD.
 
We first focus on the last term depending on the operator $\mcal{Z}^n$ (the interference contribution), as this is the most complicated one. In the interesting regime, one has
\begin{align}
    \mcal{Z}^{n,(0)}\left(\frac{x}{\xi},\ellt,\kt\right)\frac{\kgt^i}{\kg^2}\simeq \frac{-\Kt^i}{\KT^2}\mcal{Z}^{n,(0)}\left(\frac{x}{\xi},\ellt,\kt\right) \,.
\end{align}
Since the kernel in front of $\mcal{Z}^{n,(0)}$ does not depend any more upon $\ellt$, one can use the identity Eq.\,\eqref{eq:Zop-id} to perform the integral over $\ellt$. We find
\begin{align}
    -\frac{\Kt^i}{K_\perp^2}\int\frac{\rmd^2\kt}{k_\perp^2}\frac{\kt^m}{\xi}\mcal{F}_{g}^{D,(0)}\left(\frac{x}{\xi},  \kt\right)&\frac{(\Kt-\kt)^2}{\left[(1-\xi)(\Kt-\kt)^2+\xi K_\perp^2\right]}\nonumber\\
    &\times \left[\delta^{mi}-\frac{2(\Kt-\kt)^m(\Kt-\kt)^i}{(\Kt-\kt)^2}\right] \,.
\end{align}
At this stage, it is important to observe that $\kt$ is a transverse momentum scale from the target, on the same footing as $\ellt$. Hence, in the DGLAP regime where the dijet imbalance is relatively hard (and balanced by the gluon emission), $K_\perp$ is much larger than both $k_\perp$ and $\ell_\perp$. Using this, one can expand the above expression to obtain\footnote{Note that the integrand must be expanded to quadratic order in $\kt$ since the would-be linear term, as obtained by approximating $\Kt-\kt\simeq \Kt$, vanishes after the angular integration.}
\begin{align}
    -\frac{1}{K_\perp^2}\int\rmd^2\kt\,\frac{1+\xi}{\xi}\,\mcal{F}_{g}^{D,(0)}\left(\frac{x}{\xi},  \kt\right) \,.
\end{align}
After performing similar expansions of the kernels in front of the two other color correlators that appear in Eq.\,\eqref{eq:DF1R-full} (the WW and the dipole gluon TMD), one finds 
\begin{align}
         \Delta \mcal{F}_{g,\mcal{R}}^{WW}(x, K_{\perp},P_\perp^2)&\underset{K_\perp\gg \ell_\perp}{\,\simeq\,}\frac{\alpha_sN_c}{\pi^2}\frac{1}{K_\perp^2}\int\rmd^2\ellt\int_{x_*}^{1-\xi_0}\rmd \xi\left\{\frac{1}{\xi(1-\xi)}\mcal{F}_{g}^{WW,(0)}(x,\ellt)\left(\frac{x}{\xi},\ellt\right)\right.\nn*[0.2cm]
     & \left. \qquad +\frac{(1-\xi)(1+\xi^2)}{\xi}\mcal{F}_{g}^{D,(0)}\left(\frac{x}{\xi},  \ellt\right)-\frac{1+\xi}{\xi} \mcal{F}_{g}^{D,(0)}\left(\frac{x}{\xi},  \ellt\right)\right\}\,,\nn*[0.2cm]
     & \qquad =\,\frac{\alpha_s}{2\pi^2}\frac{1}{K_\perp^2}\int_{x_*}^{1-\xi_0}\rmd \xi \  P_{gg}(\xi)\,\frac{x}{\xi}G^{(0)}\left(\frac{x}{\xi},K_\perp^2\right)\,,\label{eq:rDGLAP-step}
\end{align}
where $x_*$ is a rapidity divider which separates from the B-JIMWLK evolution~\cite{Caucal:2024bae} (see also below) and $\xi_0\equiv \xi_0(\KT)$. The last line in \eqn{eq:rDGLAP-step} features the ``unregularised'' gluon-to-gluon DGLAP splitting function, defined as
\begin{align}\label{Pgg}
    P_{gg}(\xi)\equiv\frac{2N_c\left[1-\xi(1-\xi)\right]^2}{\xi(1-\xi)}\,,
\end{align}
together with LO integrated (or ``collinear'')  gluon distribution $xG^{(0)}(x,K_\perp^2)$, that has been here reconstructed from both the WW, and the dipole, gluon TMD (when $K_\perp\gg Q_s$, both definitions yield the same result):
\begin{align}\label{LOPDF}
    xG^{(0)}(x,K_\perp^2)&=\int\rmd^2\ellt \,\mcal{F}_{g}^{WW,(0)}(x,\ellt)\,\Theta(K_\perp^2-\ellt^2)\,,\nn
    &=\int\rmd^2\ellt\, \mcal{F}_{g}^{D,(0)}(x,\ellt)\,\Theta(K_\perp^2-\ellt^2)\,.
\end{align}

Within the picture of projectile evolution, the singular part $\propto 1/[\xi(1-\xi)]$ of $P_{gg}(\xi)$ fully comes from final state emissions, while the non-singular pieces are coming from both initial state emissions and interference terms~\cite{Caucal:2024bae}. But of course  Eq.\,\eqref{eq:rDGLAP-step} is most naturally interpreted in the target picture, where it describes the first step in the real DGLAP evolution of the gluon collinear PDF. The purpose of the DGLAP evolution is to resum the large transverse logarithms generated when integrating Eq.\,\eqref{eq:rDGLAP-step} over $\KT^2$ up to the hard resolution scale $\PT^2\sim Q^2$. Notice that such logarithms are also implicitly present in the LO expression for the gluon PDF in \eqn{LOPDF}: when $\ell_\perp\gg Q_s$, the gluon TMDs behave like $\mcal{F}_{g}^{WW,(0)}(x,\ellt)\propto 1/\ell_\perp^2$ (and similarly for $\mcal{F}_{g}^{D,(0)}$); hence, the LO gluon PDF $xG^{(0)}(x,K_\perp^2)$ includes a contribution enhanced by the transverse logarithm $\ln(\KT^2/Q_s^2)$.

The integral in Eq.\,\eqref{eq:rDGLAP-step} generates the logarithm $\ln(1/\xi_0)$ with $\xi_0= K_\perp/P_\perp$, via the singular behaviour of the splitting function near $\xi=1$. As explained in Sect.~\ref{sec:CSS}, this contribution truly belongs to the CSS evolution and therefore should be subtracted from Eq.\,\eqref{eq:rDGLAP-step}. This subtraction generates the plus prescription in the DGLAP splitting function, as it could have been expected. Indeed, one can write (omitting unnecessary factors and arguments)
\begin{align}
\int_{x_*}^{1-\xi_0}\frac{\rmd \xi}{1-\xi} 
\frac{x}{\xi} G^{(0)}\left(\frac{x}{\xi}\right)&=
\int_0^{1}\frac{\rmd \xi}{1-\xi} \left[
\Theta(\xi-{x_*})\frac{x}{\xi} G^{(0)}\left(\frac{x}{\xi}\right)-G^{(0)}(x)\right]
\nn*[0.2cm]
&\quad +\int_0^{1-\xi_0}\frac{\rmd \xi}{1-\xi} \,G^{(0)}(x)+\order{\xi_0}
\nn*[0.2cm]
     &= \ln\frac{1}{\xi_0}\, G^{(0)}(x, K_\perp^2) +\int_{x_*}^{1}\frac{\rmd \xi}{(1-\xi)_+} \frac{x}{\xi} G^{(0)}\left(\frac{x}{\xi}, K_\perp^2\right)+\order{\xi_0}
\,.\label{eq:plus}
\end{align}
After multiplication with $(\alpha_s N_c/\pi^2)/K_\perp^2$, the first term in the r.h.s. agrees with the  r.h.s. of \eqn{CSS0} to the accuracy of interest. Indeed, when $K_\perp\gg \ell_\perp$, the latter can be simplified as (for definiteness, we focus on the diagonal component with $m=n$)
\begin{align}
& \int\frac{\rmd^2\kgt}{\kgt^2}\,  \ln\frac{1}{\xi_0} \,\mcal{F}_g^{(0)}(x, \Kt+\kgt) = \int\frac{\rmd^2\ellt}{(\Kt-\ellt)^2}\, \ln\frac{1}{\xi_0} \,
 \mcal{F}_g^{(0)}(x, \ellt) \nn*[0.2cm]
     &\quad \simeq \frac{1}{K_\perp^2} \ln\frac{1}{\xi_0} \int\rmd^2 \ellt \, \mcal{F}_g^{(0)}(x, \ellt) \Theta(K_\perp^2-\ellt^2)
      = \frac{1}{K_\perp^2}\, \ln\frac{1}{\xi_0} \,G^{(0)}(x, K_\perp^2) \,.
 \end{align}
Hence, the r.h.s. of \eqn{eq:plus} contributes to two types of evolutions. The first term, proportional to the logarithm $\ln(1/\xi_0)$, is the real piece of an approximate (``double-logarithmic'') version of the CSS equation, in which neither the longitudinal momentum, nor the transverse momentum, are conserved at the splitting vertex~\cite{Caucal:2024bae}. The second term, involving the ``plus'' prescription, combines with the non-singular pieces of $P_{gg}(\xi)$ in \eqn{eq:rDGLAP-step} to generate the regularised version of the DGLAP splitting function, as given by \eqn{Pgg} with $1-\xi \to (1-\xi)_+$.

It remains to explain the role of the rapidity divider $x_*$ introduced in \eqn{eq:rDGLAP-step}. To the accuracy of interest, this is an arbitrary number obeying $x_*\ll 1$ and $\alpha_s\ln(1/x_*)\ll 1$. Its role is to separate between the high-energy, B-JIMWLK, evolution at $x/x_0 < \xi < x_*$ and the DGLAP evolution at larger values $x_* < \xi < 1$. This separation is needed because both evolutions develop logarithmic contributions $\propto \ln(1/\xi)$ when $\xi\ll 1$, so there is a danger of over-counting. Yet, given the different natures of the respective resummations, this over-counting refers only to the radiative corrections enhanced by a {\it double} logarithm: a small-$x$ logarithm times a transverse one. The separation scale $x_*$ is meant to avoid this over-counting. To double-logarithmic accuracy, the two evolutions are equivalent to each other, so the dependence upon $x_*$ cancels among them. The remaining dependence is, at most, of order $\alpha_s\ln(1/x_*)$, and is negligible given our above constraints on $x_*$~\cite{Caucal:2024bae}.

\subsection{The TMD $g\to gg$ splitting function for hard gluons}
\label{sec:BFKL}

Another interesting limit, which partially overlaps with the two previous ones, corresponds to the regime where the transverse momenta $K_\perp,\ell_\perp$ and $k_{g\perp}$ are much larger than $Q_s$ without assuming any hierarchy among them. We shall refer to this regime as the ``dilute" limit of the CGC, where the saturation effects can be neglected. In this limit, all the gluon TMDs share the same universal tail at large $\ell_\perp$ which can be obtained by expanding the Wilson lines (c.f. Eq.\,\eqref{eq:Wilson-line}) in the color operators to first order in $g A^-$, with $A^-=A^-_at^a$ the target gauge field. Introducing the Fourier transform of this field, defined as
\begin{align}
    \tilde\alpha(\ellt)=\int\rmd^2\bx \ e^{-i\ellt\cdot\bx}\int_{-\infty}^{\infty}\rmd x^+ A^-(x^+,\bx)\,,
\end{align}
we have
\begin{align}
    \mcal{F}_{g}^{WW,(0)}(x,\ellt)\simeq \mcal{F}_{g}^{D,(0)}(x,\ellt)\simeq \mcal{F}_g^{(0)}(x,\ellt)\equiv \frac{4\ell_\perp^2}{(2\pi)^3}\left\langle \textrm{tr}\left[  \tilde\alpha(\ellt)  \tilde\alpha^\dagger(\ellt)\right]\right\rangle_x\,,
\end{align}
at large $\ell_\perp\gg Q_s$. What is non trivial is to compute the dilute limit of the colour operator $\mathcal{Z}^n$. This can be done thanks to the following identities for the colour structure building the operator $\mathcal{Z}^n$:
\begin{align}
    \mathcal{C}_A^{a,m}(\ellt)&=-ig\ellt^m [\tilde\alpha(\ellt),t^a]+\mathcal{O}(g^2)\\
    \mathcal{C}_B^a(\ellt,\ellt')&=ig(2\pi)^2\left(\delta^{(2)}(\ellt')-\delta^{(2)}(\ellt)\right)[\tilde\alpha(\ellt+\ellt'),t^a]+\mathcal{O}(g^2)\,,
\end{align}
which imply that, for $Q_s\ll \ell_\perp,k_\perp$,
\begin{align}
    \frac{1}{k_\perp^2}\mathcal{Z}^n(x,\ellt,\kt)\simeq \left(\delta^{(2)}(\ellt-\kt)-\delta^{(2)}(\kt)\right)\frac{\ellt^n}{\ell_\perp^2}\mcal{F}_g^{(0)}(x,\ellt)\,. \label{eq:Zop-dilut}
\end{align}
Using these results, we find for the trace part of $\Delta \mcal{W}_{\mcal{R}}^{mn}$:
\begin{align}\label{eq:DF1R-full-dilute}
     &\Delta \mcal{F}_{g,\mcal{R}}^{WW}(x, K_{\perp},P_\perp^2)\underset{Q_s\ll \ell_\perp,K_\perp}{\,\simeq\,}\frac{\alpha_sN_c}{\pi^2}\int \frac{\rmd^2\ellt}{\kgt^2}\int_x^{(1-\xi_0)}\rmd \xi \ \mathcal{F}_g^{(0)}\left(\frac{x}{\xi},\ellt\right)\nn
     &\times \Bigg\{\frac{1}{\xi(1-\xi)}+\frac{(1-\xi)k_{g\perp}^2}{2\xi\ell_\perp^2}\left[\frac{\xi^2(\KT^2-\kg^2)^2}{\left((1-\xi)\KT^2+\xi\kg^2\right)^2}+
\frac{4\KT^2}{(1-\xi)\KT^2+\xi\kg^2}\bigg(1- \frac{(\kgt\cdot\Kt)^2}{\kg^2\KT^2}\bigg)\right]\nn
&-\frac{1}{\xi\ell_\perp^2}\left[(\kgt\cdot\ellt)+\frac{\KT^2(\kgt\cdot\ellt)-2(\Kt\cdot\ellt)(\kgt\cdot\Kt)}{(1-\xi)\KT^2+\xi\kg^2}\right]\Bigg\} \,,
\end{align}
with $\kgt=\ellt-\Kt$. In the regime $K_\perp\sim \ell_\perp\gg Q_s$, the real NLO correction to the WW gluon TMD is a \textit{closed} equation formulated in terms of the universal perturbative tail of the gluon TMDs given by $ \mathcal{F}_g^{(0)}(x,\ellt)$. This closed equation enables one to simplify the interpretation of Eq.\,\eqref{eq:DF1R-full-dilute} as compared to the dense non-linear regime.
The quantity inside the curly brackets (times the colour factor $2N_c$ and multiplied by an overall factor $\kgt^2$) is the TMD $g^*\to g^*g$ splitting function $P_{g^*\to g^*g}(\xi,\Kt,\kgt)$: it describes the splitting of a $t$-channel off-shell gluon with transverse momentum $\ellt$ in the dilute target into one off-shell, unpolarised, gluon with transverse momentum $\Kt$ and longitudinal momentum fraction $\xi$  and a real gluon with transverse momentum $\kgt=\ellt-\Kt$ and longitudinal momentum fraction $1-\xi$ (see also Fig.\,\ref{fig:s-vs-t-channel-picture-nlo}). Simplifying the expression in Eq.\,\eqref{eq:DF1R-full-dilute}, we have
\begin{align}
    \Delta \mcal{F}_{g,\mcal{R}}^{WW}(x, K_{\perp},P_\perp^2)\underset{\ell_\perp,K_\perp\gg Q_s}{\,\simeq\,}\frac{\alpha_s}{2\pi^2}\int \frac{\rmd^2\ellt}{\kgt^2}\int_x^{(1-\xi_0)}\rmd \xi P_{g^*\to g^*g}(\xi,\Kt,\kgt) \mathcal{F}_g^{(0)}\left(\frac{x}{\xi},\ellt\right) \,,
\end{align}
where we introduced the splitting function
\begin{align}
    &P_{g^*\to g^*g}(\xi,\Kt,\kgt) \nonumber \\
    & \equiv 2N_c\left\{\frac{1}{\xi(1-\xi)}+ \frac{\left[ (\Kt \cdot \kgt)  -  \kg^2  \right]}{\left[(1-\xi)\KT^2+\xi\kg^2\right]} + \frac{\xi(1-\xi) \kg^2 \left[ \KT^2-\kg^2 \right]^2}{2 \ell_\perp^2 \left[(1-\xi)\KT^2+\xi\kg^2\right]^2}\right\}\,. \label{eq:TMD-Pgg-unpol}
\end{align}
Note that because one of the outgoing gluon is off-shell, while the other is on-shell, the splitting function is not symmetric under $\xi\leftrightarrow1-\xi$ and $\Kt\leftrightarrow\kgt$ exchange.
The analogous TMD splitting function of an off-shell gluon into two gluons, one being linearly polarised, can be found in Appendix~\ref{app:linpol-TMD}, Eq.~\eqref{eq:TMD-Pgg-lin}.

We have explicitly checked that Eq.\,\eqref{eq:TMD-Pgg-unpol} is identical to the TMD $g\to gg$ splitting function computed in~\cite{Hentschinski:2017ayz} using $k_T$-factorisation. To further compare our result with the one obtained in~\cite{Hentschinski:2017ayz}, we have computed the TMD splitting function $\bar P$ averaged over the azimuthal angle between $\Kt$ and~$\kgt$, 
\begin{align}
    \bar P(\xi,K_\perp,\kg)=2N_c\left\{\frac{1}{\xi(1-\xi)}- \frac{\kg^2}{(1-\xi)\KT ^2+\xi\kg^2}+\frac{\xi(1-\xi)\kg^2|\KT^2-\kg^2|}{2[(1-\xi)\KT^2+\xi\kg^2]^2}\right\}\,.
\end{align}
For this azimuthally averaged splitting function, the agreement between our result and Eq.~(90) in~\cite{Hentschinski:2017ayz} is also manifest after simplifying the latter. This is a non-trivial cross-check of our calculation, in particular of the  general formula Eq.\,\eqref{eq:DF1R-full}. 

As earlier mentioned, the dilute regime discussed in this section partially overlaps with the high-energy and the collinear regimes described in the previous sections. So, we expect to recover the ``dilute'' versions of the respective results. Indeed, when $K_\perp\gg \ell_\perp$ (strong ordering in transverse momenta), Eq.\,\eqref{eq:DF1R-full-dilute} reproduces the DGLAP result in  Eq.\,\eqref{eq:rDGLAP-step}, since one can write
\begin{align}
    P_{g^*\to g^*g }(\xi,\Kt,\kgt)=P_{gg}(\xi)+\mathcal{O}\left(\frac{\ell_\perp^2}{K_\perp^2}\right)\,.
\end{align}
Furthermore, when $x\ll \xi \ll1$ (strong ordering in longitudinal momenta), Eq.\,\eqref{eq:DF1R-full-dilute} yields one (real) step in the BFKL evolution, which is the expected limit of all B-JIMWLK equations (in particular, the DMMX equation, cf. Eq.\,\eqref{eq:rDMMX}) in the dilute regime:
\begin{align}
      \Delta \mcal{F}_{g,\mcal{R}}^{WW}(x, K_{\perp},P_\perp^2)=\frac{\alpha_sN_c}{\pi^2}\ln\left(\frac{x_0}{x}\right)\int\frac{\rmd^2 \ellt}{(\ellt-\Kt)^2}\mathcal{F}_g^{(0)}\left(x,\ellt\right)\,.
\end{align}


\section{Summary and outlook}
\label{sec5}

In this paper, we have computed the leading-power, real, NLO corrections to the cross-section for the back-to-back production of quark-antiquark dijets in DIS at small $x$. We have demonstrated that these corrections preserve TMD factorisation, with the same hard factor as at tree-level and a NLO correction to the (tensorial) WW gluon TMD exhibited in \eqn{eq:DF1R-full}. This general result applies in the regime where the dijet relative momentum $P_\perp$ is the only hard transverse momentum scale in the problem, $P_\perp\gg K_\perp, Q_s$, without any special hierarchy among the ``semi-hard'' scales $K_\perp$ (the dijet imbalance) and $Q_s$ (the target saturation momentum). The photon virtuality $Q^2$ was assumed to obey $Q^2\lesssim P_\perp^2$ and all our results remain well defined in the photo-production limit $Q^2\to 0$.

By taking special limits of the general result in \eqn{eq:DF1R-full}, we have been able to recover the real parts of three different evolution equations: the DMMX equation describing the high-energy (B-JIMWLK) evolution of the WW gluon TMD, the CSS equation resumming the Sudakov double logarithms (for a discussion of the single Sudakov logarithm proportional to the one-loop coefficient $\beta_0$ of the QCD $\beta$-function in the context of CGC calculations, see \cite{Caucal:2023nci,Caucal:2024bae}), and the DGLAP equation for the gluon PDF. 

Our analysis followed a ``top-down'' strategy in which TMD factorisation at NLO and the associated quantum evolutions have been explicitly demonstrated via systematic studies of the NLO corrections to the dijet cross-section computed within the CGC EFT. Hence, our present derivations of the various evolution equations for the gluon TMD/PDF should be seen as complementary to previous ones that have instead followed a ``bottom-up'' approach --- they focused on the one-loop corrections to the operator defining the WW gluon TMD~\cite{Balitsky:2015qba,Mukherjee:2023snp}.

There are two interesting by-products of our calculation. The first one is the derivation of the B-JIMWLK evolution equation of the Weiszäcker-Williams gluon TMD (also known as DMMX equation) in momentum space, which involves the dipole gluon TMD on the right hand side. We have also discussed the linear regime where the dijet momentum imbalance and the momentum transferred by the target to the quark-antiquark-gluon system are much larger than the nucleus saturation scale. In this case, the real NLO correction is interpreted as the convolution between the UGD of the target and a TMD gluon to gluon-gluon splitting function $P_{g^*\to g^* g }$ which agrees with the expression previously obtained in~\cite{Hentschinski:2017ayz}. As discussed in appendix~\ref{app:linpol-TMD}, we also obtain for the first time the $g^*\to g^* g$ TMD splitting function when one of the outgoing gluon is linearly polarised.

Combining the results from this paper, as well as the computation of the virtual graphs done in~\cite{Caucal:2023nci,Caucal:2023fsf}, the general form of the NLO cross-section for inclusive back-to-back quark-antiquark dijet production in DIS takes the schematic form
\begin{align}
    \frac{\dif \sigma^{\gamma_{T}^* A \to q\bar{q}X}_\lo}
	{\rmd z_1
  	\rmd z_2\dif^2\Pt\dif^2\Kt } &=\left[\mathcal{H}_{\rm LO}^{mn}+\alpha_s\mathcal{H}_{\rm NLO}^{mn}\right]\mcal{W}^{mn}(x_{q\bar q},\Kt,P_\perp)+\mathcal{H}_{\rm LO}^{mn}\Delta \mcal{W}_{\mcal{R}}^{mn}(x_{q\bar q}. \Kt)\label{eq:NLOdijet-summary}
\end{align}
It is here understood that the WW gluon TMD $\mcal{W}^{mn}(x,\Kt,P_\perp)$ includes both the B-JIMWLK evolution of the operator providing the $x$ dependence of the TMD and the CSS evolution giving its $P_\perp$ dependence. The NLO hard factor $\mathcal{H}_{\rm NLO}^{mn}$ can be found in~\cite{Caucal:2023fsf}, while the new term $\Delta \mcal{W}_{\mcal{R}}^{mn}$ obtained in this paper is given by
\begin{align}\label{eq:DF1R-full-final}
     &\Delta \mcal{W}_{\mcal{R}}^{mn}(x, \Kt)=\frac{\alpha_sN_c}{\pi^2}\int\rmd^2\kgt\int_{x_\star}^{1}\rmd \xi\Bigg\{\frac{1}{\xi(1-\xi)_+}\frac{1}{\kgt^2}\mcal{W}^{mn,(0)}\left(\frac{x}{\xi},\Kt+\kgt\right)\nn
& +\frac{1-\xi}{4\xi}\frac{1}{(\Kt+\kgt)^2}\left[\frac{\xi^2(\KT^2-\kg^2)^2\delta^{mn}}{\left((1-\xi)\KT^2+\xi\kg^2\right)^2}+
\frac{4\KT^2}{(1-\xi)\KT^2+\xi\kg^2}\right.\nonumber\\
&\left.\times \bigg(\frac{\kgt^m \kgt^n}{\kg^2} +\frac{\Kt^m \Kt^n}{\KT^2}- \frac{(\kgt^m\Kt^n+\kgt^n\Kt^m) (\kgt\cdot \Kt)}{\kg^2\KT^2}\bigg)\right]\mcal{F}_{g}^{D,(0)}\left(\frac{x}{\xi},  \Kt+\kgt\right)\nn
 &+\frac{1}{2}\Bigg[\int\frac{\rmd^2\kt}{\kt^2}\frac{1}{\xi}\mcal{Z}^{n,(0)}\left(\frac{x}{\xi},\Kt+\kgt,\kt\right)\nn
&\times \frac{\kgt^i(\Kt-\kt)^2}{\kg^2\left[(1-\xi)(\Kt-\kt)^2+\xi\kg^2\right]}\left[\delta^{mi}-\frac{2(\Kt-\kt)^m(\Kt-\kt)^i}{(\Kt-\kt)^2}\right]+\overline{(m\leftrightarrow n)} \Bigg]\Bigg\} \,.
\end{align}
With respect to Eq.\,\eqref{eq:DF1R-full}, the two important modifications are \texttt{(i)} the lower bound on $\xi$ has been replaced by $x_\star$, since the small-$x$ phase space at $x/x_0< \xi < x_*$ has already been included in the high energy evolution of $\mcal{W}^{mn}$, and \texttt{(ii)}  the upper bound has been replaced by $1$ (while adding the plus prescription on the would-be singular factor $1/(1-\xi)$), since $\mcal{W}^{mn}$ is also assumed to include the CSS evolution. Note that the tree-level operators on the right hand side of Eq.\,\eqref{eq:DF1R-full-final} can be promoted to the evolved ones, although strictly speaking, such a manipulation should be demonstrated by a two-loop calculation as it goes beyond the NLO perturbative accuracy of our calculation.

The additional term $\Delta \mcal{W}_{\mcal{R}}^{mn}(x, \Kt)$ is a finite $\mathcal{O}(\alpha_s)$ correction which can be understood as a NLO correction to the WW gluon TMD operator itself. In other words, we conjecture that one can simply replace $\mcal{W}^{mn}(x_{q\bar q},\Kt,P_\perp^2)$ by $\mcal{W}^{mn}(x_{q\bar q},\Kt,P_\perp^2)+\Delta \mcal{W}_{\mcal{R}}^{mn}(x,\Kt )$ in the first term of the right hand side of Eq.\,\eqref{eq:NLOdijet-summary} to include a subset of the two-loops $\mathcal{O}(\alpha_s^2)$ impact factor from this process coming from the product between the NLO hard factor and $\Delta \mcal{W}_{\mcal{R}}^{mn}(x, \Kt)$. 

We leave for future work the numerical evaluation of this finite piece, but we anticipate it to be challenging if no further approximation is considered because of the presence of the operator $\mathcal{Z}(x,\ktone,\kttwo)$ defined in Eq.\,\eqref{eq:Zop-def} which involves a derivative of the quadrupole operator at small $x$. In the future, we also intend to carry out a similar analysis for forward prompt photon-jet production in proton-nucleus collisions (which at tree-level is known to admit TMD factorisation in terms of the dipole gluon TMD~\cite{Dominguez:2011wm}).

\paragraph{Acknowledgements.} We would like to thank Marcos Guerrero Morales, Cyrille Marquet, Dionysis Triantafyllopoulos, Raju Venugopalan and Shu-Yi Wei for stimulating discussions. We are grateful for the support of the Saturated Glue (SURGE) and the  Quark-Gluon Tomography (QGT) Topical Theory Collaborations, funded by the U.S. Department of Energy, Office of Science, Office of Nuclear Physics, and in part by the Office of Science of the U.S. Department of Energy under Contract No. DE-AC02-05CH11231. P. C. thanks the EIC theory institute at BNL for support during the initial stages of this work. P.C. and F.S. also thank the Aspen Center for Physics, which is supported by National Science Foundation grant PHY-2210452, where this work was completed in part. F.S. was supported in part by a grant from the Alfred P. Sloan Foundation (G-2024-22395).  Figures were created with JaxoDraw~\cite{Binosi:2003yf}.

\appendix

\section{Derivation of the amplitude for inclusive back-to-back $q\bar q$ production plus a soft gluon in $\gamma^*$-nucleus collisions}
\label{app:amplitude-qqbarg}

In this Appendix, we provide the key steps in deriving Eqs.\,\eqref{PsigBfull}-\eqref{PsigAfull} in the main text. 

\subsection{Amplitudes: general results in momentum space}
\label{sub:amplitude-general}

\begin{figure}
    \centering
    \includegraphics[width=0.8\linewidth]{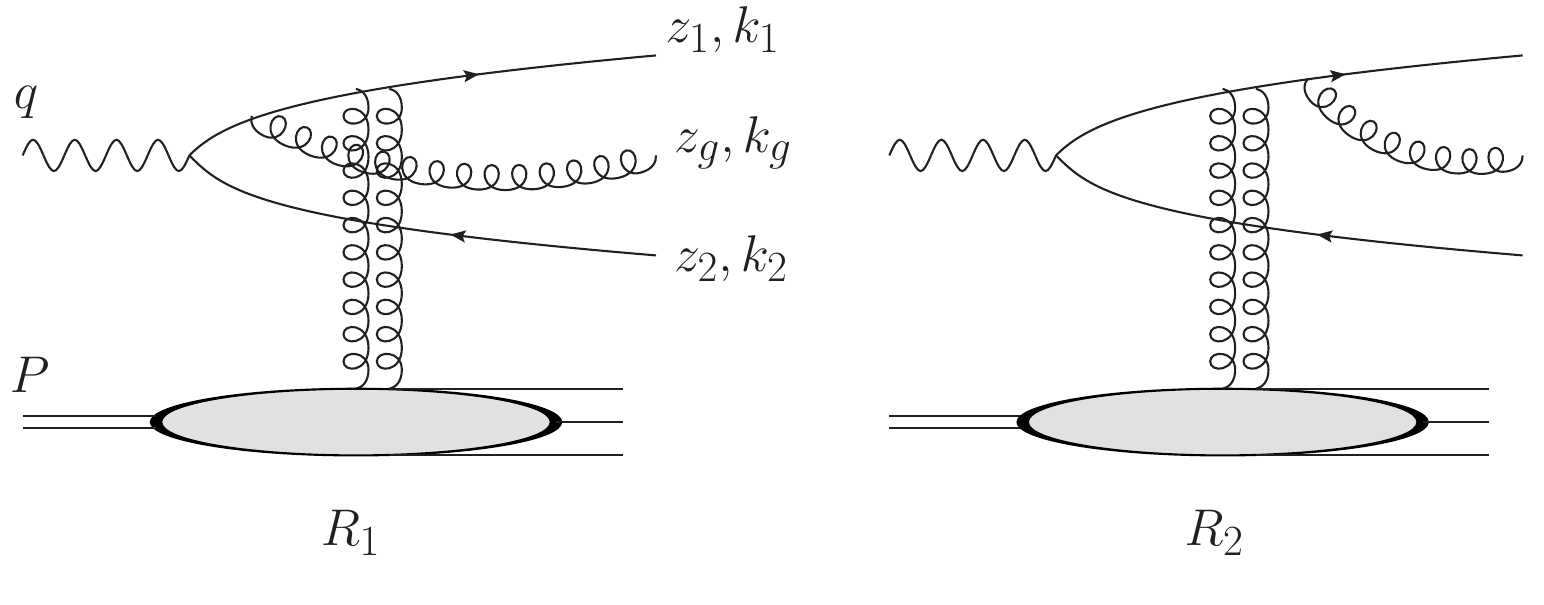}
    \caption{Real NLO amplitudes, labelled $R_1$ and $R_2$, for the $\gamma^*A\to q\bar qX$ process. The gluon is emitted either before (left) or after (right) the shock-wave, represented by the vertical gluons. The amplitudes $R_3$ and $R_4$ where the gluon is emitted by the anti-quark are not shown.}
    \label{fig:CGC-real-graphs}
\end{figure}

The real amplitudes for gluon emission by the $q\bar q$ dipole have been computed in~\cite{Ayala:2017rmh,Iancu:2018hwa,Caucal:2021ent,Bergabo:2023wed} for the dijet process in DIS. The corresponding Feynman graphs are displayed in Fig.\,\ref{fig:CGC-real-graphs}.
In order to perform the expansions needed for the calculation of two hard $q\bar q$ jets plus one soft gluon jet, it will be more convenient to express the expansion in momentum space. Hence, in this section we present the amplitudes in momentum space. These expressions are valid in general kinematics at small $x$. The detailed derivation of the momentum space formula from the coordinate space expressions displayed in~\cite{Caucal:2021ent} can be found in the supplemental material of~\cite{Caucal:2025xxh}. We provide the expressions for the amplitudes for gluon emission from the quark. The amplitudes from emission from antiquark can be easily obtained by quark-antiquark interchange.

We label $\qtone$ and $\qttwo$ the internal (loop) momenta transferred from the target to the quark and antiquark respectively. The momentum transfer to the gluon is then fixed by momentum conservation. 

These choices of momenta will be more convenient when carrying out the power expansion in the hard scale $\textrm{max}(P_\perp,Q)$. The typical values of these momenta are indeed of the order of the soft scales $q_{1\perp}, q_{2\perp} \sim Q_s, K_\perp$. We can then write the amplitude for gluon emission before the shock-wave as:
\begin{align}
    \mathcal{M}_{R_1, \sigma\bar\sigma, i \Bar{i}}^{\lambda, \Bar{\lambda},a}(k_1,k_2,k_g) = \int \frac{\rmd^2 \qtone}{(2\pi)^2} \int \frac{\rmd^2 \qttwo}{(2\pi)^2} \Ccal^a_{R_1, i \bar{i}}(\qtone,\qttwo) \Jcal_{R_1, \sigma\bar\sigma}^{\lambda, \bar \lambda}(\qtone,\qttwo) \,,
    \label{eq:amplitude-emission-before-SW-quark}
\end{align}
where the colour factor is
\begin{align}
    \Ccal^{a}_{R_1, i \bar{i}}(\qtone,\qttwo) =& \int \rmd^2 \xt \rmd^2 \yt \rmd^2 \zt e^{-i \qtone  \cdot \xt} e^{-i \qttwo \cdot \yt} e^{-i (\ktone + \kttwo + \kgt - \qtone - \qttwo) \cdot \zt} \nonumber \\
    & \times \left[ V(\xt) V^\dagger(\zt) t^a V(\zt) V^\dagger(\yt) - t^a  \right]_{i \bar{i}} \,.
\end{align}
In these expressions, $\sigma,\bar \sigma$, $i,\bar i$ refer to the helicity and colour indices of the quark and antiquark respectively, $\bar\lambda$ and $a$ are the polarisation and colour labels of the radiated gluon and $\lambda$ is the polarisation index of the incoming virtual photon. The perturbative factor for longitudinally polarized photons is
\begin{align}
    &\Jcal_{R_1,\sigma\bar\sigma}^{\lambda=0,\bar \lambda}(\qtone,\qttwo)  =  -\frac{e e_f g}{z_g} \frac{4   z_2 (1- z_2)Q}{z_2(1-z_2)Q^2  + (\qttwo - \kttwo)^2}\nonumber  \\
    &\times \frac{\et^{\bar\lambda*,n} \bar{u}_\sigma(k_1) \left\{ \left( \qtone -\ktone + \frac{z_1}{1- z_2} (\qttwo - \kttwo) \right)_m  \left[\left( 1 + \frac{z_g}{2z_1}\right) \delta^{mn} - \frac{z_g}{2z_1} \omega^{mn}\right] \right \} \gamma^+ v_{\bar{\sigma}}(k_2)}{ Q^2 + \frac{(\qttwo - \kttwo)^2}{z_2} + \frac{(\qtone - \ktone)^2  }{z_1} + \frac{(\qtone + \qttwo - \ktone -\kttwo)^2}{z_g}} \,,
\end{align}
and for transversely polarized photons is
\begin{align}
    &\Jcal_{R_1,\sigma\bar\sigma}^{\lambda=\pm 1,\bar \lambda}(\qtone,\qttwo)  = - \frac{e e_f g \et^{\bar\lambda*,n} \et^{\lambda,i} }{\left[ Q^2 + \frac{(\qttwo - \kttwo)^2}{z_2} + \frac{(\qtone - \ktone)^2  }{z_1} + \frac{(\qtone + \qttwo - \ktone -\kttwo)^2}{z_g}  \right]} \nonumber \\
    &  \times \bar{u}_\sigma(k_1) \Bigg[ \frac{2  \left\{ \left( \qtone -\ktone + \frac{z_1}{1- z_2} (\qttwo - \kttwo) \right)_m  \left[\left( 1 + \frac{z_g}{2z_1}\right) \delta^{mn} - \frac{z_g}{2z_1} \omega^{mn}\right]    \right \}}{z_g \left[Q^2z_2(1-z_2)  + (\qttwo - \kttwo)^2 \right]} \nonumber \\
    & \times \left\{ (\qttwo - \kttwo)_j \left[ \delta^{ij} (1- 2 z_2) - \omega^{ij}\right] +  \gamma^i   \right\}   +  \frac{ \gamma^n  \gamma^i  }{(1-z_2)} \Bigg] \gamma^+ v_{\bar{\sigma}}(k_2) \,.
\end{align}
In these expressions, we have introduced the commutator of transverse gamma matrices $\omega^{ij} \equiv \frac{1}{2} \left[\gamma^i, \gamma^j \right]$. 

The amplitude for a gluon emission from the quark after the shock-wave can be cast in the similar form. However, since there is only one internal (loop) momentum, the momentum space convolution only involves one momentum transfer, which we choose to be the momentum transfer to the anti-quark $\qttwo$:
\begin{align}
    \mathcal{M}_{R_2, \sigma\bar\sigma, i \Bar{i}}^{\lambda, \Bar{\lambda},a}(k_1,k_2,k_g) = \int \frac{\rmd^2 \qttwo}{(2\pi)^2} \Ccal^a_{R_2,i \bar{i}}(\qttwo) \Jcal^{\lambda, \bar \lambda}_{R_2, \sigma\bar\sigma}(\qttwo) \,,
\end{align}
where the colour factor is
\begin{align}
    \Ccal^a_{R_2, i \bar{i}}(\qttwo) &= \int \rmd^2 \xt \rmd^2 \yt e^{-i (\ktone + \kttwo + \kgt -\qttwo) \cdot \xt} e^{-i \qttwo \cdot \yt}  t^a \left[V(\xt) V^\dagger(\yt)  - 1 \right]_{i \bar{i}} \,.
    \label{eq:amplitude-emission-after-SW-quark}
\end{align}
The perturbative factor for longitudinally polarized photons is
\begin{align}
    \Jcal^{\lambda=0, \bar \lambda}_{R_2,\sigma\bar\sigma}(\qttwo)&=\frac{ 4e e_f  g (1-z_2) z_2  Q}{(1-z_2) z_2  Q^2  + (\qttwo - \kttwo)^2 }\nonumber\\
    &\times \frac{\et^{\bar\lambda*,n} \bar{u}_\sigma(k_1) \left\{(z_1 \kgt- z_g \ktone)_m \left[\delta^{mn}\left(1 + \frac{z_g}{2z_1}\right) - \frac{z_g}{2 z_1} \omega^{mn}\right]  \right\}   \gamma^+ v_{\bar{\sigma}}(k_2)}{\frac{1}{z_1} (z_1 \kgt - z_g \ktone)^2} \,, 
\end{align}
and for transversely polarized photons
\begin{align}
    \Jcal^{\lambda=\pm 1, \bar \lambda}_{R_2,\sigma\bar\sigma}(\qttwo) & =- e e_f  g  \frac{2 \et^{\bar\lambda*,n}  \et^{\lambda,i} \bar{u}_\sigma(k_1) \left\{(z_1 \kgt- z_g \ktone)_m \left[\delta^{mn}\left(1 + \frac{z_g}{2z_1}\right) - \frac{z_g}{2 z_1} \omega^{mn}\right] \right\} }{\left[ (1-z_2) z_2  Q^2  + (\qttwo - \kttwo)^2 \right] \left[\frac{1}{z_1} (z_1 \kgt - z_g \ktone)^2 \right]}  \nonumber \\
    & \times \left\{ (\qttwo-\kttwo)_{j} \left[\delta^{ij} (2z_2 - 1) + \omega^{ij}\right]  \right\}  \gamma^+ v_{\bar{\sigma}}(k_2) \,.
\end{align}

\subsection{Amplitudes: back-to-back limit}

We turn now to the soft gluon limit of the amplitudes written in the previous subsection. As in the main text, we define
\begin{align}
    \Pt = \frac{z_2 \ktone - z_1 \kttwo}{z_1 + z_2} \,,\quad  
    \Kt = \ktone + \kttwo \,, \quad
    \ellt = \ktone + \kttwo + \kgt \,,
\end{align}
and we consider the kinematic limit:
\begin{align}
     Q_s^2, K_\perp^2 \ll P_\perp^2,\, Q^2 \,, \quad
     z_1,z_2 \sim \mathcal{O}(1) \,.
\end{align}
The conditions $K_\perp^2, Q_s^2 \ll P_\perp^2, Q^2$ implie that
\begin{align}
    \qtone^2, \qttwo^2  \ll P_\perp^2,\, Q^2 \,.
\end{align}
Note that this is due to our choice of momenta definition $\qtone$ and $\qttwo$, namely, the momenta correspond to the momenta transfer from the shock-wave to the quark and antiquark respectively, which is controlled by the intrinsic saturation scale $Q_s$.
Observe also that these conditions imply  $\kgt^2 \ll \Pt^2,\, Q^2 $. Due to the structure of the ``light-cone" energy denominators in the amplitudes for gluon emissions before the shockwave, the integral over $z_g$ will be dominated by small values of $z_g$
\begin{align}
    z_{g} \sim \frac{\mathrm{max}(\Kt^2,Q_s^2)}{\mathrm{max}(\Pt^2,Q^2)} \ll 1 \,.
\end{align}
Therefore, in this limit, the gluon has small transverse momentum $k_{g\perp}$ as compared to the transverse momentum of the quark and antiquark $k_{1\perp} \sim k_{2\perp} \sim P_\perp$, as well as small longitudinal momentum fraction $z_{g} \ll z_1, z_2$.

\subsubsection*{Expansion of perturbative factors $R_1$ and $R_3$}

Introducing the expansion parameter $\lambda$
\begin{align}
    \lambda^2 \sim  \frac{\mathrm{max}(\Kt^2,Q_s^2)}{\mathrm{max}(\Pt^2,Q^2)} \sim z_g \,.
\end{align}
We must expand the perturbative factor up to linear order in $\lambda$ since the zeroth order will cancel when combining contributions of gluon emission from quark and antiquark. This cancellation is a consequence of the suppression of gluon emission from a small dipole, i.e. the gluon cannot resolve the dipole separation.

Let us begin by expanding the light-cone energy denominators for gluon emission from quark and before the shock-wave:
\begin{align}
    &\left[ Q^2 + \frac{\left(\qttwo  +\Pt - \frac{z_2 \Kt}{z_1 + z_2}\right)^2}{z_2} + \frac{\left(\qtone - \Pt - \frac{z_1 \Kt}{z_1 + z_2} \right)^2  }{z_1} + \frac{(\qtone + \qttwo - \Kt)^2}{z_g}  \right] \nonumber \\
    & \simeq \left[ Q^2 + \frac{\Pt^2 }{z_1 z_2}  + \frac{(\qtone + \qttwo - \Kt)^2}{z_g}  \right] + \frac{2 \Pt \cdot ( z_1 \qttwo - z_2 \qtone )}{z_1 z_2} + \cdots \,, \label{eq:denominator1_NLO}
\end{align}
and
\begin{align}
    &\left[z_2(1-z_2) Q^2  + \left(\qttwo +\Pt - \frac{z_2 \Kt}{z_1 + z_2} \right)^2 \right]  \simeq  \left[z_1 z_2 Q^2  + \Pt^2 \right] + 2 \Pt \cdot \left(\qttwo - z_2 \Kt \right) \,.
\end{align}
Next, let us expand the numerator. We use
\begin{align}
    \ktone + \frac{z_1}{1-z_2} \kttwo&\simeq \Kt + \frac{z_g \Pt}{z_1} \,, 
\end{align}
where in this approximation, we have dropped terms quadratic in $\lambda$. Then we have
\begin{align}
    \qtone  -\ktone + \frac{z_1}{1- z_2} (\qttwo - \kttwo) \simeq \qtone + \qttwo  - \Kt - \frac{z_g}{z_1} \Pt \,.
\end{align}
The gluon emission from antiquark is obtained easily following the rules given in section~\ref{sub:amplitude-general}.

\paragraph{Longitudinally polarised virtual photon.}
Inserting these expansions, we find that the contribution from gluon emission before the shock-wave to the  perturbative factor is 
\begin{align}
    &\Jcal_{R_1,\sigma\bar\sigma}^{\lambda=0,\bar \lambda} + \Jcal_{R_3,\sigma\bar\sigma}^{\lambda=0,\bar \lambda}  = -\frac{e e_f g 4  z_1 z_2 Q   \et^{\bar\lambda*,n} \bar{u}_\sigma(k_1)\gamma^+ v_{\bar{\sigma}}(k_2) }{z_g \left\{\left[ Q^2 + \frac{\Pt^2 }{z_1 z_2}  + \frac{(\qtone + \qttwo - \Kt)^2}{z_g}  \right] + \frac{2 \Pt \cdot ( z_1 \qttwo - z_2 \qtone )}{z_1 z_2}  \right\}}    \\
    & \times \left\{ \frac{  \left( \qtone + \qttwo  - \Kt - \frac{z_g}{z_1} \Pt \right)^n     }{ \left[z_1 z_2 Q^2  + \Pt^2 \right] + 2 \Pt \cdot \left(\qttwo - z_2 \Kt \right) } - \frac{  \left( \qtone + \qttwo  - \Kt + \frac{z_g}{z_2} \Pt \right)^n     }{ \left[z_1 z_2 Q^2  + \Pt^2 \right] - 2 \Pt \cdot \left(\qtone - z_1 \Kt \right) } \right \} \,.
\end{align}
Let's focus on the term in the curly bracket, as anticipated the leading contribution cancels when combining quark and antiquark contributions, we have 
\begin{align}
    &\frac{  \left( \qtone + \qttwo  - \Kt - \frac{z_g}{z_1} \Pt \right)^n     }{ \left[z_1 z_2 Q^2  + \Pt^2 \right] + 2 \Pt \cdot \left(\qttwo - z_2 \Kt \right) } - \frac{  \left( \qtone + \qttwo  - \Kt + \frac{z_g}{z_2} \Pt \right)^n     }{ \left[z_1 z_2 Q^2  + \Pt^2 \right] - 2 \Pt \cdot \left(\qtone - z_1 \Kt \right) }  \nonumber \\
   & \simeq - \frac{z_g}{z_1 z_2} \frac{\Pt_m}{\left[z_1 z_2 Q^2  + \Pt^2 \right] } \left[   \delta^{mn}  +  \frac{2   \left(\qtone + \qttwo -  \Kt \right)^m \left( \qtone + \qttwo  - \Kt \right)^n }{z_g \left[ Q^2 +  \frac{\Pt^2}{z_1 z_2}\right]}\right] \,.
\end{align}
Since this piece starts next-to-leading order in $\lambda$; it is sufficient to keep the leading order piee in Eq.\,\eqref{eq:denominator1_NLO}, thus we have
\begin{align}
    &\Jcal_{R_1,\sigma\bar\sigma}^{\lambda=0,\bar \lambda}(\qtone,\qttwo) + \Jcal_{R_3,\sigma\bar\sigma}^{\lambda=0,\bar \lambda}(\qtone,\qttwo) =   \frac{ 8 e e_f g  z_1 z_2 Q \bar{u}_\sigma(k_1) \gamma^+ v_{\bar{\sigma}}(k_2) \Pt_m  \et^{\bar\lambda*,n}}{\left[z_1 z_2 Q^2  + \Pt^2 \right]^2} \nonumber \\
    & \times \frac{\left[ \left(\qtone + \qttwo -  \Kt \right)^m \left( \qtone + \qttwo  - \Kt \right)^n + z_g \left( Q^2 + \frac{\Pt^2 }{z_1 z_2} \right)  \frac{\delta^{mn}}{2}\right] }{ \left[ (\qtone + \qttwo - \Kt)^2  + z_g \left(Q^2 + \frac{\Pt^2 }{z_1 z_2} \right) \right]}  \,.
\end{align}
Observe that the perturbative factor only depends on the sum of momenta $\qtone+\qttwo$ which is the \textit{total} momentum transfer to the quark anti-quark pair.

\paragraph{Transversely polarised virtual photon}

The case $\lambda=\pm1$ can be worked out in a similar fashion, and it yields
\begin{align}
    & \Jcal_{R_1,\sigma\bar\sigma}^{\lambda=\pm 1,\bar \lambda}(\qtone,\qttwo) + \Jcal_{R_3,\sigma\bar\sigma}^{\lambda=\pm 1,\bar \lambda}(\qtone,\qttwo) \nonumber \\
    &=
    -\frac{2 e e_f g \et^{\bar\lambda*,n} \et^{\lambda,i} }{  \left[ z_1 z_2 Q^2 + \Pt^2   \right] } \bar{u}_\sigma(k_1) \left\{ \left[ \delta_{mj} -  \frac{2  \Pt_{m} \Pt_j  }{\left[z_1 z_2 Q^2  + \Pt^2 \right]} \right] \left[ \delta^{ij} (z_1 - z_2) - \omega^{ij}\right] \right\} \gamma^+ v_{\bar{\sigma}}(k_2) \nonumber \\
    & \times \frac{\left[ \left(\qtone + \qttwo - \Kt \right)^m  \left( \qtone + \qttwo  - \Kt \right)^n + z_g \left( Q^2 + \frac{\Pt^2  }{z_1 z_2} \right) \frac{\delta^{mn}}{2}   \right]}{\left[ (\qtone + \qttwo - \Kt)^2 + z_g \left( Q^2 + \frac{\Pt^2 }{z_1 z_2} \right) \right]} \,.
\end{align}

\subsubsection*{Expansion of perturbative factors $R_2$ and $R_4$}
We now consider the contributions from gluon emission after the shock-wave. While their energy denominators do not strongly constrain $z_g$, we focus on gluon emissions outside of the jet cone; hence, we shall approximate (c.f. discussion Sec.\,\ref{sec:phase-space-constraints})
\begin{align}
    \left(\kgt-\frac{z_g}{z_{i}}\Pt\right)^2\simeq \kgt^2 \,,
\end{align}
for $i=1,2$.

\paragraph{Longitudinally polarised virtual photon.}
The leading term in the expansion:
\begin{align}
    \Jcal_{R_2, \sigma\bar\sigma}^{\lambda=0, \bar \lambda}(\qttwo)= e e_f  g  \frac{4Q z_1 z_2 \et^{\bar\lambda*,n} \bar{u}_\sigma(k_1)   \gamma^+ v_{\bar{\sigma}}(k_2)}{\left[ z_1 z_2  Q^2  + \Pt^2 \right]}   \frac{\kgt^n}{\kgt^2} + \dots 
\end{align}
is independent of $\qttwo$. Since
\begin{align}
    \int \frac{\rmd^2 \qttwo}{(2\pi)^2} \Ccal^a_{R_2,i \bar{i}}(\qttwo) & = \int \rmd^2 \xt e^{-i (\ktone + \kttwo + \kgt) \cdot \xt}  t^a \left[V(\xt) V^\dagger(\xt)  - \mathbbm{1} \right]_{i \bar{i}} \nonumber \\
    & = 0 \,,
\end{align}
due to unitarity, then the leading term of the expansion of the perturbative factor results in a vanishing contribution. The physical reason behind this cancellation is that the leading order contribution corresponds dipole of zero size which will not interact with the shock-wave due to color transparency.

Let us compute the next-to-leading power expansion. It is sufficient to consider the contribution linear in $\qttwo$. Using
\begin{align}
    &\frac{1}{\left[z_2(1-z_2) Q^2  + \left(\qttwo +\Pt - \frac{z_2 \Kt}{z_1 + z_2} \right)^2 \right]}    & \simeq  \frac{1}{\left[z_1 z_2 Q^2  + \Pt^2 \right]}\left[1 - \frac{2 \Pt \cdot \qttwo}{\left[z_1 z_2 Q^2  + \Pt^2 \right]} \right] \,,
\end{align}
and keeping only the $\qttwo$ dependent term, we find
\begin{align}
    \Jcal_{R_2, \sigma\bar\sigma}^{\lambda=0, \bar \lambda}(\qttwo) &= - e e_f  g  \frac{8Q z_1 z_2 \et^{\bar\lambda*,n} \bar{u}_\sigma(k_1)   \gamma^+ v_{\bar{\sigma}}(k_2)}{\left[z_1 z_2 Q^2  + \Pt^2 \right]} \frac{ \Pt_{m}}{\left[z_1 z_2 Q^2  + \Pt^2 \right]} \frac{\qttwo^m \kgt^n}{\kgt^2} + \dots \,.
\end{align}
Similarly, we have
\begin{align}
    \Jcal_{R_4, \sigma\bar\sigma}^{\lambda=0, \bar \lambda}(\qttwo) &= e e_f  g  \frac{8Q z_1 z_2 \et^{\bar\lambda*,n} \bar{u}_\sigma(k_1)   \gamma^+ v_{\bar{\sigma}}(k_2)}{\left[z_1 z_2 Q^2  + \Pt^2 \right]} \frac{ \Pt_{m}}{\left[z_1 z_2 Q^2  + \Pt^2 \right]} \frac{\qttwo^m \kgt^n}{\kgt^2} + \dots\,,\nonumber \\
    & = -\Jcal_{R_2, \sigma\bar\sigma}^{\lambda=0, \bar \lambda}(\qttwo) \,,
\end{align}
where we have used the quark-antiquark symmetry $\qttwo \to \ktone + \kttwo + \kgt -\qttwo$ (and keeping only the term linear in $\qttwo$), $\Pt \to -\Pt$, and an overall minus sign in front of the perturbative factor.

\paragraph{Transversely polarised virtual photon.}

Once again, we expand to next-to-leading power but only keep the $\qttwo$ dependent term
\begin{align}
    \Jcal_{R_2, \sigma\bar\sigma}^{\lambda=\pm 1, \bar \lambda}(\qttwo)
    &= \frac{2e e_f  g \et^{\bar\lambda*,n} \et^{\lambda,i}}{\left[z_1 z_2 Q^2  + \Pt^2 \right]} \frac{\qttwo^m \kgt^n}{\kgt^2}    \nonumber \\
    & \times \bar{u}_\sigma(k_1)    \left\{ \left[\delta_{jm} -\frac{2 \Pt_j \Pt_m}{\left[z_1 z_2 Q^2  + \Pt^2 \right]}\right]\left[\delta^{ij} (z_1-z_2) - \omega^{ij}\right] \right\}   \gamma^+ v_{\bar{\sigma}}(k_2) \,.
\end{align}
Similarly, we find for the antiquark contribution
\begin{align}
    \Jcal_{R_4, \sigma\bar\sigma}^{\lambda=\pm 1, \bar \lambda}(\qttwo) = -\Jcal_{R_2, \sigma\bar\sigma}^{\lambda=\pm 1, \bar \lambda}(\qttwo) \,.
\end{align}

\subsubsection*{Simplifying the colour correlators}

First, we recap our results for the perturbative factors. For gluon emission before the shock-wave, either by the quark or the antiquark, the sum of the two amplitudes can be expressed as
\begin{align}
    & \Jcal_{R_1,\sigma\bar\sigma}^{\lambda,\bar \lambda} + \Jcal_{R_3,\sigma\bar\sigma}^{\lambda,\bar \lambda} = \Ical^{\lambda}_{m,\sigma\bar\sigma} (\Pt, Q^2, z_1) \et^{\bar\lambda*,n} \Gamma^{mn}(\qtone +\qttwo-\Kt)  \,,
\end{align}
while for gluon emissions after the shock-wave, they read
\begin{align}
    \Jcal_{R_2,\sigma\bar\sigma}^{\lambda,\bar \lambda}(\qtone,\qttwo) &= -\Ical^{\lambda}_{m,\sigma\bar\sigma} (\Pt, Q^2, z_1) \et^{\bar\lambda*,n} \frac{\qttwo^m \kgt^n}{\kgt^2} \,,\\
    \Jcal_{R_4,\sigma\bar\sigma}^{\lambda}(\qtone,\qttwo) &=\Ical^{\lambda, \bar \lambda}_{m,\sigma\bar\sigma} (\Pt, Q^2, z_1) \et^{\bar\lambda*,n} \frac{\qttwo^m \kgt^n}{\kgt^2} \,,
\end{align}
for an emission by the quark and the antiquark respectively.
Here we defined the shorthand notations for the gluon emission kernel by the effective gluon-gluon dipole,
\begin{align}
    \Gamma^{mn}(\qt) = \frac{\left[ \qt^m  \qt^n + z_g \left( Q^2 + \frac{\Pt^2  }{z_1 z_2} \right) \frac{\delta^{mn}}{2}   \right]}{\left[ \qt^2 + z_g \left( Q^2 + \frac{\Pt^2 }{z_1 z_2} \right) \right]} \,. 
    \label{eq:splitting-before-SW}
\end{align}
The hard factors at the amplitude level are
\begin{align}
    &\Ical^{\lambda=0}_{m,\sigma\bar\sigma}(\Pt, Q^2, z_1) =  \frac{ 8 e e_f g    z_1 z_2 Q \bar{u}_\sigma(k_1) \gamma^+ v_{\bar{\sigma}}(k_2) \Pt_m }{\left[z_1 z_2 Q^2  + \Pt^2 \right]^2} \,,  \label{eq:soft-gluon-amplitude-hard-factor-longitudinal}\\
    & \Ical^{\lambda=\pm 1}_{m,\sigma\bar\sigma}(\Pt, Q^2, z_1) = -\frac{2 e e_f g  \et^{\lambda,i} }{  \left[ z_1 z_2 Q^2 + \Pt^2   \right] } \nonumber \\
    & \times \bar{u}_\sigma(k_1) \left\{ \left[ \delta_{mj} -  \frac{2  \Pt_{m} \Pt_j  }{\left[z_1 z_2 Q^2  + \Pt^2 \right]} \right] \left[ \delta^{ij} (z_1 - z_2) - \omega^{ij}\right] \right\} \gamma^+ v_{\bar{\sigma}}(k_2) \,. \label{eq:soft-gluon-amplitude-hard-factor-transverse}
\end{align}

We now wish to simplify the colour structure built from Wilson lines. Let us consider the contribution coming from the emission before the shock-wave. It is convenient to introduce the change of variables:
\begin{align}
    \kt &= \qtone + \qttwo \,, \quad\Qt = \frac{1}{2}(\qtone -\qttwo) \,,
\end{align}
such that the integration variable $\kt$ physically represents the total momentum transferred to the quark \textit{and} the antiquark at the level of the amplitude for gluon emission before the shock-wave.

Since the perturbative factor is independent of $\Qt$ then we can directly integrate the colour factor:
\begin{align}
\int \frac{\rmd^2 \Qt}{(2\pi)^2}& \Ccal^a_{R_1, i \bar{i}}\left(\Qt + \frac{\kt}{2},-\Qt + \frac{\kt}{2}\right)\nonumber\\
& = \int \rmd^2 \xt  \ \rmd^2 \zt \ e^{-i \kt \cdot \xt} e^{-i (\ellt - \kt) \cdot \zt} \left[ V(\xt) V^\dagger(\zt) t^a V(\zt) V^\dagger(\xt) - t^a  \right]_{i \bar{i}} \,.
\end{align}
Then we have
\begin{align}
    \mathcal{M}_{R_1+R_3, \sigma\bar\sigma, i \Bar{i}}^{\lambda, \Bar{\lambda},a} = \Ical^{\lambda}_{m,\sigma\bar\sigma} (\Pt, Q^2, z_1) \et^{\bar\lambda*,n} \int \frac{\rmd^2 \kt}{(2\pi)^2} \Ccal^a_{B,i \bar{i}}(\kt,\ellt-\kt)  \Gamma^{mn}(\kt-\Kt) \,,\label{eq:MR1-R3-app-final}
\end{align}
where defined the colour factor
\begin{align}
    \Ccal^a_{B,i \bar{i}}(\ellt,\ellt')  = \int \rmd^2 \xt \  \rmd^2 \zt \ e^{-i \ellt \cdot \xt} e^{-i \ellt' \cdot \zt} \left[ V(\xt) V^\dagger(\zt) t^a V(\zt) V^\dagger(\xt) - t^a  \right]_{i \bar{i}} \,.
\end{align}
After carrying out the spin-helicity contraction in the hard factor and using the definition of the adjoint representation to express $\mathcal{C}_B$ in terms of Wilson lines in the adjoint representation, Eqs.\,\eqref{eq:MR1-R3-app-final} is identical to Eq.\,\eqref{PsigBfull} provided in the main text.

We finally move on to the contribution coming from emissions after the shock-wave:
\begin{align}
    \mathcal{M}_{R_2, \sigma\bar\sigma, i \Bar{i}}^{\lambda, \Bar{\lambda},a} =  \Ical^{\lambda}_{m,\sigma\bar\sigma} (\Pt, Q^2, z_1)  \et^{\bar\lambda*,n} \frac{\kgt^n}{\kgt^2} \Ccal^{a,m}_{R_2,i\bar{i}}(\Kt+\kgt) \,, 
\end{align}
where
\begin{align}
    \Ccal^{a,m}_{R_2,i \bar{i}}(\Kt+\kgt)   & = \int \frac{\rmd^2 \qttwo}{(2\pi)^2} (-\qttwo)^m \Ccal^a_{R_2,i \bar{i}}(\qttwo) \nonumber \\
   & = i \int \rmd^2 \xt \ e^{-i (\Kt + \kgt) \cdot \xt}  \left[  t^a V(\xt) \partial^m_\perp V^\dagger(\xt) \right]_{i \bar{i}} \,.
\end{align}
Similarly, exploiting the quark anti-quark symmetry, the sum of amplitudes for emission after the shock-wave is
\begin{align}
    \mathcal{M}_{R_2+R_4, \sigma\bar\sigma, i \Bar{i}}^{\lambda, \Bar{\lambda},a} =  \Ical^{\lambda}_{m} (\Pt, Q^2, z_1) \et^{\bar\lambda*,n} \frac{\kgt^n}{\kgt^2} \Ccal^{a,m}_{A,i \bar{i}}(\Kt+\kgt) \,,\label{eq:MR2-R4-app-final}
\end{align}
where
\begin{align}
    \Ccal^{a,m}_{A,i \bar{i}}(\ellt) = \int \rmd^2 \xt \ e^{-i \ellt \cdot \xt}  \left[t^a , i \left( V(\xt) \partial^m_\perp V^\dagger(\xt) \right) \right]_{i \bar{i}} \,.
\end{align}
Here again, Eq.\,\eqref{eq:MR2-R4-app-final} is identical to Eq.\,\eqref{PsigAfull} used in the main text after spin-helicity contraction.

\section{Linearly polarised Weiszäcker-Williams gluons}
\label{app:linpol-TMD}

In this Appendix, we discuss the linearly polarised WW gluon TMD at NLO in the Colour Glass Condensate.

\paragraph{TMD and DGLAP $g^*\to g^*_Lg$ splitting function.} We first give the result for the TMD $g^*_L\to g^*_Lg$ splitting function in the dense regime (without any approximation on the relative order of magnitude between $K_\perp, \ell_\perp$ and $Q_s$) in the case where the parent gluon is linearly polarised. The correction $\Delta \mathcal{H}_g^{WW,(0)}$ to the linearly polarised WW gluon TMD $\mathcal{H}_{g,\mcal{R}}^{WW}(x,K_\perp,P_\perp^2)$ can be obtained from $\Delta\mcal{W}^{mn}_{\mcal{R}}(x,K_\perp,P_\perp^2)$ using
\begin{align}
    \Delta \mathcal{H}_{g,\mcal{R}}^{WW}(x,\KT,P_\perp^2)=\Delta\mcal{W}^{mn}_{\mcal{R}}(x,\Kt)\left[\frac{2\Kt^m\Kt^n}{K_\perp^2}-\delta^{mn}\right]
\end{align}
and plugging the expression given by Eq.\,\eqref{eq:DF1R-full} for $\Delta\mcal{W}^{mn}_{\mcal{R}}(x,\ellt)$, we find that the contribution depending on the dipole gluon TMD vanishes, such that the remaining contribution reads
\begin{align}\label{eq:deltaHWW-full}
      &\Delta \mathcal{H}_{g,\mcal{R}}^{WW}=\frac{\alpha_s N_c}{\pi^2}\int\rmd^2\ellt\int_x^{1-\xi_0}\rmd\xi\Bigg\{\frac{1}{\xi(1-\xi)\kg^2}\left[\frac{2(\Kt\cdot\ellt)^2}{\KT^2\ell_\perp^2}-1\right]\mcal{H}_g^{WW,(0)}\left(\frac{x}{\xi},\ellt\right)\nn
      &+\int\frac{\rmd^2\kt}{\xi k_\perp^2}\frac{\kgt^i (\Kt-\kt)^2 \left[\delta^{mi}-\frac{2(\Kt-\kt)^m (\Kt-\kt)^i}{(\Kt-\kt)^2}\right]}{\kg^2\left[(1-\xi)(\Kt-\kt)^2+\xi\kg^2\right]}\left[\frac{2\Kt^m\Kt^n}{\KT^2}-\delta^{mn}\right]\mcal{Z}^{n,(0)}\left(\frac{x}{\xi},\ellt,\kt\right) \,,
\end{align}
where $\kgt=\ellt-\Kt$.
This expression is the full $g\to gg$ real splitting vertex for an incoming linearly polarised gluon in the non-linear regime. 

Finally, we use Eq.\,\eqref{eq:Zop-dilut} and the fact that the unpolarised and linearly polarised WW gluon TMD share the same dilute tail, i.e,
\begin{align}
    \mathcal{H}_g^{WW,(0)}(x,\ellt)\underset{\ell_\perp\gg Q_s}\simeq\mcal{H}_g^{(0)}(x,\ellt)=\mcal{F}_g^{(0)}(x,\ellt)\,,
\end{align}
which physically means that at large $\ell_\perp$, all incoming gluons are linearly polarised. Then, the dilute limit $\ell_\perp,K_\perp\gg Q_s$ of equation Eq.\,\eqref{eq:deltaHWW-full} reads
\begin{align}
      &\Delta \mathcal{H}_{g,\mcal{R}}^{WW}(x,\KT,P_\perp^2)=\frac{\alpha_s N_c}{\pi^2}\int\rmd^2\ellt\int_x^{1-\xi_0}\rmd\xi \ \mcal{H}_g^{(0)}\left(\frac{x}{\xi},\ellt\right)\Bigg\{\frac{1}{(1-\xi)\kg^2}\left[\frac{2(\Kt\cdot\ellt)^2}{\KT^2\ell_\perp^2}-1\right]\nn
      &+\frac{(\ellt\cdot\Kt)}{\xi \kg^2\ell_\perp^2}+\frac{\KT^2(\kgt\cdot\ellt)}{\xi\ell_\perp^2\kg^2\left[(1-\xi)\KT^2+\xi\kg^2\right]}\Bigg\} \,.
\end{align}
One reads from this expression --- by multiplying the quantity inside the curly brackets by $2N_c\kg^2$ --- the TMD splitting $P_{g^*\to g^*_Lg}(\xi,\Kt,\kgt)$ of a linearly polarised gluon with transverse momentum $\ellt=\Kt+\kgt$ into two gluon with transverse momenta $\Kt$ and $\kgt$ sharing a longitudinal momentum fraction $\xi,(1-\xi)$ respectively:
\begin{align}\label{eq:TMD-Pgg-lin}
    & P_{g^*\to g_L^*g}(\xi,\Kt,\kgt) \nonumber \\
    & \equiv 2N_c\Bigg\{\frac{1}{\xi} + \frac{1}{(1-\xi)}\left[\frac{2(\Kt\cdot\ellt)^2}{\KT^2\ell_\perp^2}-1\right] + \frac{\left[\KT^2 -\kg^2\right]\left[\kg^2 - (\Kt \cdot \kgt)\right]}{\ell_\perp^2\left[(1-\xi)\KT^2+\xi\kg^2\right]}\Bigg\}\,.
\end{align}
Last but not least, by expanding this result when the hierarchy of scales $K_\perp\gg \ell_\perp$ holds, we recover the collinear splitting function of a linearly polarised gluon, discussed in~\cite{Sun:2011iw,Caucal:2024bae}:
\begin{align}
      &\Delta \mathcal{H}_{g,\mcal{R}}^{WW}(x,\KT,P_\perp^2)=\frac{\alpha_s}{2\pi^2}\frac{1}{\KT^2}\int_x^{1-\xi_0}\rmd\xi\frac{2N_c(1-\xi)}{\xi}\frac{x}{\xi}G^{(0)}\left(\frac{x}{\xi},K_\perp^2\right)
\end{align}
or equivalently,
\begin{align}
    P_{g^*\to g_L^*g}(\xi,\Kt,\kgt)=\frac{2N_c(1-\xi)}{\xi}+\mathcal{O}\left(\frac{\ell_\perp^2}{K_\perp^2}\right)
\end{align}

\paragraph{CSS evolution of the linearly polarised WW gluon TMD.} The evolution of the linearly polarised WW gluon TMD can be obtained from the generalisation of Eq.\,\eqref{CSSKT} to the full tensor structure $\mcal{W}^{mn}$ of the WW gluon TMD, namely
\begin{align}
	\label{CSSKT-Wmn}
 \frac{\del \mcal{W}^{mn}(x,\Kt, P_\perp^2)}{\del \ln P_\perp^2}\,=\frac{\alpha_sN_c}{2\pi}\int\frac{\rmd^2\kgt}{\pi\kgt^2}\, 
  &\Big[\mcal{W}^{mn}(x, \Kt  +\kgt, P_\perp^2) \nonumber\\
  &- \Theta(P_\perp^2-\kg^2)\mcal{W}^{mn}(x,\Kt, P_\perp^2)\Big]. \end{align}
In particular, the momentum space CSS equation is obtained from this equation by contracting the tensor structure with $2K_\perp^mK_\perp^n/\KT^2-\delta^{mn}$ such that
\begin{align}
	\label{CSSKT-Hg}
 \frac{\del \mcal{H}_g(x,\Kt, P_\perp^2)}{\del \ln P_\perp^2}\,=\frac{\alpha_sN_c}{2\pi^2}\int\frac{\rmd^2\kgt}{\kgt^2}\, 
  &\Bigg\{\left(\frac{2[(\Kt+\kgt)\cdot\Kt]^2}{(\Kt+\kgt)^2K_\perp^2}-1\right)\mcal{H}_g(x,\Kt +\kgt, P_\perp^2) \nonumber\\
  &- \Theta(P_\perp^2-\kg^2)\mcal{H}_g(x,\Kt  , P_\perp^2)\Bigg\}. \end{align}
Note that the real part of this equation exactly matches the first line in Eq.\,\eqref{eq:deltaHWW-full} after taking the derivative with respect to $\ln(P_\perp^2)$, as it should. 
We now want to obtain the corresponding CSS equation in coordinate space. Instead of using Eq.\,\eqref{CSSKT-Hg} and the relation between the linearly polarised WW gluon TMD in coordinate space vs momentum space, namely
\begin{align}
     \tilde{\mcal{H}}_g(x, \rt, P_\perp^2)=\int\frac{\rmd^2\Kt}{(2\pi)^2}e^{i\Kt\cdot\rt}\left[\frac{2(\Kt\cdot\rt)^2}{K_\perp^2r_\perp^2}-1\right]\mcal{H}_g(x, \Kt  , P_\perp^2)\,,
\end{align}
 it is actually more convenient to first take the Fourier transform of Eq.\,\eqref{CSSKT-Wmn} and then perform the contraction with the tensor $2r_\perp^m r_\perp^n/r_\perp^2-\delta^{mn}$ which isolates the coordinate space expression of the linearly polarised WW gluon TMD. Taking the Fourier transform of Eq.\,\eqref{CSSKT-Wmn} and with the help of the identity
\begin{align}
    \int\frac{\rmd^2\kgt}{\kgt^2}\left[e^{-i\kgt\cdot\rt}-\Theta(P_\perp^2-\kg^2)\right]&=-\pi\ln^2\left(\frac{P_\perp^2r_\perp^2}{c_0^2}\right)
\end{align}
one shows that the WW gluon operator in coordinate space satisfies
\begin{align}
	\label{CSSbT-Wmn}
 \frac{\del \tilde{\mcal{W}}^{mn}(x, \rt, P_\perp^2)}{\del \ln P_\perp^2}&\,=\,-\frac{\alpha_sN_c}{2\pi}\left[\ln\frac{r^2P_\perp^2}{c_0}\right]
 \, \tilde{\mcal{W}}^{mn}(x, \rt, P_\perp^2)\,, \end{align}
 such that, since
 \begin{align}
    \tilde{\mcal{H}}_g(x, \rt, P_\perp^2)&\equiv \tilde{\mcal{W}}^{mn}(x, \rt, P_\perp^2)\left[\frac{2 r_\perp^mr_\perp^n}{r_\perp^2}-\delta^{mn}\right]\,,
\end{align}
the linearly polarised WW gluon TMD in coordinate space satisfies the same equation as the unpolarised one,
 \begin{align}
     	\label{CSSbT-h}
 \frac{\del \tilde{\mcal{H}}_g(x, \rt, P_\perp^2)}{\del \ln P_\perp^2}&\,=\,-\frac{\alpha_sN_c}{2\pi}\left[\ln\frac{r^2P_\perp^2}{c_0}\right]
 \, \tilde{\mcal{H}}_g(x, \rt, P_\perp^2)\,.
 \end{align}

\section{DMMX equation: from coordinate to momentum space}
\label{app:DMMX}

The DMMX equation has been obtained in \cite{Dominguez:2011br}. The real gluon emission contribution to the DMMX (rDMMX) can be identified by replacing the full kernel by:
\begin{align}
    \frac{(\bb-\bbb)^2}{(\zt-\bb)^2 (\bbb-\zt)^2} \to \frac{2(\zt-\bb) \cdot (\bbb-\zt)}{(\zt-\bb)^2 (\bbb-\zt)^2}  \,,
\end{align}
which corresponds to real emissions only. Then the rDMMX equation reads:
\begin{align}
    & \frac{\partial}{\partial \ln(1/x)} \left \langle \tr\left[ V(\bb) (\partial^m V^\dagger(\bb)) V(\bbb) (\partial^n V^\dagger(\bbb))  \right] \right\rangle_{x}  \nonumber \\
    &= -\frac{\alpha_s N_c}{\pi^2} \int \der^2 \zt \frac{(\zt-\bb) \cdot (\bbb-\zt)}{(\zt-\bb)^2 (\bbb-\zt)^2} \left \langle \tr\left[ V(\bb) (\partial^m V^\dagger(\bb)) V(\bbb) (\partial^n V^\dagger(\bbb))  \right] \right\rangle_{x}  \nonumber \\
    & -\frac{\alpha_s N_c}{2\pi^2} \int \der^2 \zt  \left[\frac{\partial}{\partial \bb^m} \frac{(\zt-\bb) \cdot (\bbb-\zt)}{(\zt-\bb)^2 (\bbb-\zt)^2} \right] \nonumber \\
    &\times \frac{1}{N_c} \Bigg\{ \left \langle \tr\left[ V(\zt) V^\dagger(\bb) V(\bbb) (\partial^n V^\dagger(\bbb))\right] \tr\left[ V(\bb) V^\dagger(\zt) \right] \right\rangle_{x}  - ( \bb \leftrightarrow \zt) \Bigg\} \nonumber \\
    & -\frac{\alpha_s N_c}{2\pi^2} \int \der^2 \zt  \left[\frac{\partial}{\partial \bbb^n} \frac{(\zt-\bb) \cdot (\bbb-\zt)}{(\zt-\bb)^2 (\bbb-\zt)^2} \right] \nonumber \\
    & \times \frac{1}{N_c} \Bigg\{ \left \langle \tr\left[ V(\bb) (\partial^m V^\dagger(\bb)) V(\zt) V^\dagger(\bbb) \right]  \tr\left[ V(\bbb) V^\dagger(\zt) \right] \right\rangle_{x}  - ( \bbb \leftrightarrow \zt) \Bigg\} \nonumber \\
    & -\frac{\alpha_s N_c}{2\pi^2} \int \der^2 \zt \left[\frac{\partial}{\partial \bb^m}  \frac{\partial}{\partial \bbb^n} \frac{(\zt-\bb) \cdot (\bbb-\zt)}{(\zt-\bb)^2 (\bbb-\zt)^2} \right] \frac{1}{N_c} \Bigg \{  \left \langle \tr\left[V(\zt) V^\dagger(\bbb) \right]  \tr\left[V(\bbb) V^\dagger(\zt) \right] \right\rangle_{x}   \nonumber \\
    & +  \left. \Big \langle \tr\left[V(\zt) V^\dagger(\bb) \right]\tr\left[V(\bb) V^\dagger(\zt) \right] \right\rangle_{x}  -  \left \langle \tr\left[V(\bb) V^\dagger(\bbb) \right] \tr\left[V(\bbb) V^\dagger(\bb) \right] \right\rangle_{x}  -N_c^2 \Bigg\} \,. 
    \label{eq:DMMX-coordinate-space}
\end{align}
The goal of this appendix is to demonstrate that this equation is equivalent to the momentum space version presented in Eq.\,\eqref{eq:rDMMX}. Leveraging the relations between correlators in the fundamental and adjoint representations:
\begin{align}
    & \tr\left[ V(\zt) V^\dagger(\bb) V(\bbb) (\partial^n V^\dagger(\bbb))\right] \tr\left[ V(\bb) V^\dagger(\zt) \right] - ( \bb \leftrightarrow \zt) \nonumber \\
    & = - \Tr[U(\zt) U^\dagger(\bb) (\partial^n U(\bbb)) U^\dagger(\bbb)] \,, \\
    &\tr\left[ V(\bb) (\partial^m V^\dagger(\bb)) V(\zt) V^\dagger(\bbb) \right] \tr\left[ V(\bbb) V^\dagger(\zt) \right] - ( \bbb \leftrightarrow \zt) \nonumber \\
    & = - \Tr[U(\zt) U^\dagger(\bbb) (\partial^m U(\bb)) U^\dagger(\bb)] \,, \\
    &\Bigg \{  \tr\left[V(\zt) V^\dagger(\bbb) \right] \tr\left[V(\bbb) V^\dagger(\zt) \right]  +  \tr\left[V(\zt) V^\dagger(\bb) \right] \tr\left[V(\bb) V^\dagger(\zt) \right] \nonumber \\
    & - \tr\left[V(\bb) V^\dagger(\bbb) \right] \tr\left[V(\bbb) V^\dagger(\bb) \right] - N_c^2 \Bigg\} \nonumber \\
    & =  \Bigg \{ \Tr\left[U(\zt) U^\dagger(\bbb) \right]   + \Tr\left[U(\bb) U^\dagger(\zt) \right]  -  \Tr\left[U(\bb) U^\dagger(\bbb) \right]  -(N_c^2-1) \Bigg\} \,,
\end{align}
we arrive at an alternative form of the rDMMX equation:
\begin{align}
    & \frac{\partial}{\partial \ln(1/x)} \left \langle \tr\left[ V(\bb) (\partial^m V^\dagger(\bb)) V(\bbb) (\partial^n V^\dagger(\bbb))  \right] \right \rangle_x \\
    &= -\frac{\alpha_s N_c}{\pi^2} \int \der^2 \zt \frac{ (\zt-\bb) \cdot (\bbb-\zt)}{(\zt-\bb)^2 (\bbb-\zt)^2} \left \langle \tr\left[ V(\bb) (\partial^m V^\dagger(\bb)) V(\bbb) (\partial^n V^\dagger(\bbb))  \right] \right \rangle_x \nonumber \\
    & +\frac{\alpha_s N_c}{2\pi^2} \int \der^2 \zt  \left[\frac{\partial}{\partial \bb^m} \frac{ (\zt-\bb) \cdot (\bbb-\zt)}{(\zt-\bb)^2 (\bbb-\zt)^2} \right]  \frac{1}{N_c} \left \langle \Tr[U(\zt) U^\dagger(\bb) (\partial^n U(\bbb)) U^\dagger(\bbb)] \right \rangle_x \nonumber \\
    &  +\frac{\alpha_s N_c}{2\pi^2} \int \der^2 \zt  \left[\frac{\partial}{\partial \bbb^n} \frac{ (\zt-\bb) \cdot (\bbb-\zt)}{(\zt-\bb)^2 (\bbb-\zt)^2} \right] \frac{1}{N_c} \left \langle \Tr[U(\zt) U^\dagger(\bbb) (\partial^m U(\bb)) U^\dagger(\bb)] \right \rangle_x \nonumber \\
    & -\frac{\alpha_s N_c}{2\pi^2} \int \der^2 \zt \left[\frac{\partial}{\partial \bb^m}  \frac{\partial}{\partial \bbb^n} \frac{ (\zt-\bb) \cdot (\bbb-\zt)}{(\zt-\bb)^2 (\bbb-\zt)^2} \right] \frac{1}{N_c} \Bigg \{ \left \langle \Tr\left[U(\zt) U^\dagger(\bbb) \right] \right \rangle_x  \nonumber \\
     & + \left \langle \Tr\left[U(\bb) U^\dagger(\zt) \right] \right \rangle_x -  \left \langle \Tr\left[U(\bb) U^\dagger(\bbb) \right] \right \rangle_x  -(N_c^2-1) \Bigg\} \,.
\end{align}
To show the equivalence to Eq.\,\eqref{eq:rDMMX}, we shall write this equation in momentum space. Taking the Fourier transform on both sides (and multiplying by $-1/(2\pi^2 \alpha_s)$), we find
\begin{align}
    & \frac{\partial}{\partial \ln(1/x)} \mcal{W}^{mn,(0)}(x, \ellt)= \frac{\alpha_s N_c}{\pi^2} \int \frac{\der^2 \bb \der^2 \bbb}{(2\pi)^2}  e^{-i \Kt \cdot (\bb-\bbb)} \nonumber \\
    & \times \Bigg\{ \int \der^2 \zt \frac{ (\zt-\bb) \cdot (\bbb-\zt)}{(\zt-\bb)^2 (\bbb-\zt)^2} \frac{1}{2\pi^2 \alpha_s} \left \langle \tr\left[ V(\bb) (\partial^m V^\dagger(\bb)) V(\bbb) (\partial^n V^\dagger(\bbb))  \right] \right \rangle_x \nonumber \\
    & -  \int \der^2 \zt  \left[\frac{\partial}{\partial \bb^m} \frac{ (\zt-\bb) \cdot (\bbb-\zt)}{(\zt-\bb)^2 (\bbb-\zt)^2} \right] \frac{1}{4\pi^2 \alpha_s} \frac{1}{N_c} \left \langle \Tr[U(\zt) U^\dagger(\bb) (\partial^n U(\bbb)) U^\dagger(\bbb)] \right \rangle_x  \nonumber \\
    &  - \int \der^2 \zt  \left[\frac{\partial}{\partial \bbb^n} \frac{ (\zt-\bb) \cdot (\bbb-\zt)}{(\zt-\bb)^2 (\bbb-\zt)^2} \right] \frac{1}{4\pi^2 \alpha_s} \left \langle \frac{1}{N_c} \Tr[U(\zt) U^\dagger(\bbb) (\partial^m U(\bb)) U^\dagger(\bb)] \right \rangle_x \nonumber \\
    & + \int \der^2 \zt \left[\frac{\partial}{\partial \bb^m}  \frac{\partial}{\partial \bbb^n} \frac{ (\zt-\bb) \cdot (\bbb-\zt)}{(\zt-\bb)^2 (\bbb-\zt)^2} \right] \frac{1}{4\pi^2 \alpha_s}  \frac{1}{N_c} \Big \{ \left \langle \Tr\left[U(\zt) U^\dagger(\bbb) \right] \right \rangle_x \nonumber \\
     & + \left \langle \Tr\left[U(\bb) U^\dagger(\zt) \right] \right \rangle_x -  \left \langle \Tr\left[U(\bb) U^\dagger(\bbb) \right] \right \rangle_x  -(N_c^2-1) \Big\} \Bigg\} \,.
     \label{eq:rDMMX2}
\end{align}
Next, we use the Fourier transform of the kernels
\begin{align}
     \frac{(\zt-\at)^m}{(\zt-\at)^2} = -i\int \frac{\der^2 \ellt}{(2\pi)} e^{i \ellt\cdot (\zt-\at)}  \frac{\ellt^m}{\ellt^2} \,,
\end{align}
\begin{align}
     \frac{\partial}{\partial \at^k} \frac{(\zt-\at)^m}{(\zt-\at)^2} = -\int \frac{\der^2 \ellt}{(2\pi)} e^{i \ellt\cdot (\zt-\at)}  \frac{\ellt^m \ellt^k}{\ellt^2} \,,
\end{align}
as well as the correlators (by Fourier inverting the definitions in Eqs.\,\eqref{dGWW},\eqref{eq:gluon-dipole},\eqref{eq:Zop-def})
\begin{align}
    \frac{S_\perp}{2\pi^2 \alpha_s} \left \langle \tr\left[ V(\bb) ( \partial^m V^\dagger(\bb)) V(\bbb) (\partial^n V^\dagger(\bbb))  \right] \right \rangle_x  &= -\int \der^2 \ellt e^{i \ellt \cdot (\bb-\bbb)} \mcal{W}^{mn,(0)}(x, \ellt) \,, \\
    \frac{S_\perp}{4\pi^2 \alpha_s}  \frac{1}{N_c} \left \langle \Tr\left[ U(\bb) U^\dagger(\bbb) \right] \right \rangle_x &= \int \der^2 \ellt e^{-i \ellt\cdot (\bb-\bbb)} \frac{\mcal{F}_{g}^{D,(0)}(\ellt)}{\ellt^2} \,, \\
    \frac{S_\perp}{4\pi^2 \alpha_s}  \frac{i}{N_c} \left \langle \Tr[U(\zt) U^\dagger(\bb) ( \partial^n U(\bbb)) U^\dagger(\bbb)] \right \rangle_x &= \int \der^2 \ellt\int \der^2 \ellt' e^{i\ellt\cdot (\zt-\bbb)} e^{i \ellt' \cdot (\bb-\zt)} \frac{\mcal{Z}^{n,(0)}(x,\ellt,\ellt’)}{\ellt'^2} \,,
\end{align}
where $S_\perp$ is an overall transverse area factor due to translational invariance. 

Using these identities, we find
\begin{align}
    &   \int \frac{\der^2 \bb \der^2 \bbb}{(2\pi)^2}  e^{-i \Kt \cdot (\bb-\bbb)}  \int \der^2 \zt \frac{(\zt-\bb) \cdot (\bbb-\zt)}{(\zt-\bb)^2 (\bbb-\zt)^2} \frac{1}{2\pi^2 \alpha_s} \left \langle \tr\left[ V(\bb) (\partial^m V^\dagger(\bb)) V(\bbb) (\partial^n V^\dagger(\bbb))  \right] \right \rangle_x \nonumber \\
    & =  \int \der^2 \ellt  \frac{\mcal{W}^{mn,(0)}(x, \ellt)}{(\ellt-\Kt)^2} \,,  
\end{align}
\begin{align}
    &  \int \frac{\der^2 \bb \der^2 \bbb}{(2\pi)^2}  e^{-i \Kt \cdot (\bb-\bbb)}  \int \der^2 \zt \left[\frac{\partial}{\partial \bb^m}  \frac{\partial}{\partial \bbb^n} \frac{ (\zt-\bb) \cdot (\bbb-\zt)}{(\zt-\bb)^2 (\bbb-\zt)^2} \right] \nonumber \\
    & \times \frac{1}{4\pi^2 \alpha_s} \frac{1}{N_c} \Bigg \{ \left \langle \Tr\left[U(\zt) U^\dagger(\bbb) \right]  \right \rangle_x + \left \langle  \Tr\left[U(\bb) U^\dagger(\zt) \right] \right \rangle_x -  \left \langle  \Tr\left[U(\bb) U^\dagger(\bbb) \right] \right \rangle_x  -(N_c^2-1) \Bigg\} \nonumber \\
    & =    \int \der^2 \ellt\left[\frac{(\ellt-\Kt)^m (\ellt-\Kt)^k}{(\ellt-\Kt)^2} - \frac{\Kt^m \Kt^k}{\Kt^2} \right]\left[\frac{(\ellt-\Kt)^n (\ellt-\Kt)^k}{(\ellt-\Kt)^2} - \frac{\Kt^n \Kt^k}{\Kt^2} \right] \frac{\mcal{F}_{g}^{D,(0)}(\ellt)}{\ellt^2} \,,
\end{align}
\begin{align}
    & \int \frac{\der^2 \bb \der^2 \bbb}{(2\pi)^2}  e^{-i \Kt \cdot (\bb-\bbb)} \int \der^2 \zt  \left[\frac{\partial}{\partial \bbb^n} \frac{ (\zt-\bb) \cdot (\bbb-\zt)}{(\zt-\bb)^2 (\bbb-\zt)^2} \right] \frac{1}{4\pi^2 \alpha_s} \frac{1}{N_c} \left \langle \Tr[U(\zt) U^\dagger(\bbb) (\partial^m U(\bb)) U^\dagger(\bb)] \right \rangle_x \nonumber \\
    & =  \int \der^2 \ellt\int \der^2 \ellt' \frac{(\ellt-\Kt)^k}{(\ellt-\Kt)^2}    \frac{(\Kt-\ellt')^n (\Kt-\ellt')^k}{(\Kt-\ellt')^2}    \frac{\mcal{Z}^{m,(0)}(x,\ellt,\ellt’)}{\ellt'^2} \,.
\end{align}
Combining these three relations, one can readily show that Eq.\,\eqref{eq:rDMMX2}, is equivalent to the momentum space version obtained in Eq.\,\eqref{eq:rDMMX}.

\bibliographystyle{utcaps}
\bibliography{refs}

\end{document}